\documentclass[transmag]{IEEEtran}
\usepackage{latexsym}
\usepackage{graphicx}
\usepackage{amsfonts,amssymb,amsmath}
\usepackage{cite}
\usepackage{booktabs}
\usepackage{multirow}
\usepackage{hyperref}
\def\BibTeX{{\rm B\kern-.05em{\sc i\kern-.025em b}\kern-.08em T\kern-.1667em\lower.7ex\hbox{E}\kern-.125emX}}
\begin{document}

\title{A Survey on Energy Optimization Techniques in UAV-Based Cellular Networks: From Conventional to Machine Learning Approaches}

\author{Attai Ibrahim Abubakar, Iftikhar Ahmad, Kenechi G. Omeke, Metin Ozturk, Cihat Ozturk, Ali Makine Abdel-Salam, \\
Michael S. Mollel, Qammer H. Abbasi, Sajjad Hussain, Muhammad Ali Imran
\thanks{This work is supported in parts by  EPSRC IAA award is EP/R511705/1.}
\thanks{Attai Ibrahim Abubakar, Iftikhar Ahmad, Kenechi G. Omeke, Qammer H. Abbasi, Sajjad Hussain, and Muhammad Ali Imran are with Communication, sensing and imaging (CSI) research group, James Watt School of Engineering, University of Glasgow, United Kingdom.}
\thanks{Metin Ozturk, Cihat Ozturk, Ali Makine Abdel-Salam are with 
Faculty of Engineering and Natural Sciences, Ankara Yıldırım Beyazıt University, Turkey.}
\thanks{Michael S. Mollel is with Nelson Mandela African Institution of Science and Technology (NM-AIST), Arusha, Tanzania.} 
\thanks{Corresponding Email: a.abubakar.1@research.gla.ac.uk.}}

\IEEEtitleabstractindextext{\begin{abstract}
Wireless communication networks have been witnessing an unprecedented demand due to the increasing number of connected devices and emerging bandwidth-hungry applications.
Albeit many competent technologies for capacity enhancement purposes, such as millimeter wave communications and network densification, there is still room and need for further capacity enhancement in wireless communication networks, especially for the cases of unusual people gatherings, such as sport competitions, musical concerts, etc.
Unmanned aerial vehicles~(UAVs) have been identified as one of the promising options to enhance the capacity due to their easy implementation, pop-up fashion operation, and cost-effective nature.
The main idea is to deploy base stations on UAVs and operate them as flying base stations, thereby bringing additional capacity to where it is needed.
However, because the UAVs mostly have limited energy storage, their energy consumption must be optimized to increase flight time.
In this survey, we investigate different energy optimization techniques with a top-level classification in terms of the optimization algorithm employed---conventional and machine learning~(ML).
Such classification helps understand the state-of-the-art and the current trend in terms of methodology.
In this regard, various optimization techniques are identified from the related literature, and they are presented under the above-mentioned classes of employed optimization methods.
In addition, for the purpose of completeness, we include a brief tutorial on the optimization methods and power supply and charging mechanisms of UAVs.
Moreover, novel concepts, such as reflective intelligent surfaces and landing spot optimization, are also covered to capture the latest trend in the literature.

\end{abstract}

}

\maketitle

\section{INTRODUCTION}

\IEEEPARstart{T}{he} fifth generation of mobile communication networks~(5G) is no longer a future technology; it has become a reality as we have been witnessing its initial deployments around the globe.
It has come with some rigorous requirements as well as promising scenarios and great expectations.
In fact, the expectations mainly originate from those requirements; for example, 5G New Radio~(5G NR) includes ultra-reliable low-latency communications~(URLLC) scenario\footnote{Enhanced mobile broadband~(eMBB) and massive machine-type communications~(mMTC) are the other two scenarios mentioned in 5G NR.} with stringent reliability and delay requirements, and it is expected to enable various use-cases, such as remote surgery, industrial automation, etc.~\cite{urllc-survey}.
Therefore, the expectations of the previously mentioned spectacular technologies are based on the requirements, which determine the level of advancement in  mobile communications technology.
Moreover, the expectations also make the requirements stiffer, thus, there is a loop around the expectations~(demands) and the toughness in the requirements~(advancement in the technology).

In addition to the requirements, scenarios, and potential use-cases, there is another challenge that 5G networks are projected to face: Internet of things~(IoT) networks.
Billions of IoT devices will be connected to the Internet over the next few years, and a considerable amount of them will be cellular IoT with a significant portion of broadband IoT~\cite{5g-iot-survey,ericsson_mr_nov21}.
There is an important takeaway from this: IoT not only dramatically increases the number of connected devices but also significantly surges the bandwidth demand.
Given that many bandwidth-hungry applications, including metaverse technology, 4K video streaming, tactile Internet, etc., are already tied together with 5G networks, the extra bandwidth demand coming from broadband IoT will put the existing challenges to a higher level.
However, 5G already has some tools to alleviate such pressure in terms of capacity enhancement, such as millimeter wave~(mmWave) communications, network densification, and massive multiple-input multiple-output~(MIMO) technology, to name a few.

Since 5G is a reality, studies on the sixth generation of mobile communication networks~(6G) has already started and moved quite forward~\cite{6g-survey-1, 6g-survey-2, 6g-survey-3, 6g-survey-4}.
Similar to the sharp advancement from the legacy networks to 5G, 6G is expected to be a paradigm shift in terms of network approaches, as mentioned in a recent report in~\cite{abiresearch}.
More specifically, 6G is envisioned to shift the focus from conventional requirements including delay, data rate, reliability to  global coverage, CO$_2$ emission, spectral efficiency, etc.~\cite{abiresearch}.
However, this does not mean that 6G will not improve the data rate, latency, and reliability of the existing networks, because it is expected that 6G will enhance the aforementioned conventional requirements significantly~\cite{6g-survey-1}, which indeed makes it more challenging for 6G networks.
That is why, for instance, the terahertz~(THz) spectrum is considered for 6G in order to open up more bandwidth for capacity enhancement and increased data rate purposes~\cite{6g-survey-1,6g-survey-2,6g-survey-3,6g-survey-4}.

The picture that we have been trying to frame so far is that there is and will be a need for capacity enhancement in the current and upcoming generations of mobile communication networks.
As mentioned earlier in this section, although each generation involves various concepts to make them immune to the rise in the capacity demand, there is always a need for higher capacity provision.
Furthermore, during scenarios involving unusual gathering of people, such as music concerts, sports competitions, fairs, etc., that seldom happens, a crowd cluster around a hotspot (i.e., the location of the event), increasing the level of challenge for the existing cellular network infrastructures, which are typically not designed for such rare events.
If not for special occasions, in a usual scenario, there are spatio-temporal changes in the traffic loads of base stations~(BSs), so that, for a particular BS, the traffic load changes dynamically over the time of a day, and days of a week~\cite{spatio-temporal,metin-vfa,nature-data, abubakar2022lightweight}.
This poses an interesting phenomenon that BSs do not always operate at full load; rather the peak times of a day, for example, are limited to a certain period.

In this regard, a pop-up solution would be required for such kinds of scenarios, and unmanned aerial vehicle (UAV)-assisted wireless networking has been introduced as a viable option for pop-up networking~\cite{mozaffari2019tutorial}. 
The idea here is that small cell~(SC) BSs are deployed on UAVs (called UAV-BSs), and they are sent to the hotspots for capacity enhancement purposes as long as there is a need, providing pop-up connectivity.
With that, not only the users are satisfied as they secure connectivity, but also the mobile network operators are relaxed as no fixed infrastructure deployment is necessary.

In addition to the capacity enhancement purposes, UAV-BSs can also be used in emergency scenarios, including earthquakes, floods, etc., where the fixed infrastructure is either fully or partially damaged and down~\cite{paulo-uav}.
Since the end-to-end connectivity cannot be provided in such scenarios, UAV-BSs come as helping hands; such that they are sent to affected regions and operate either as standalone BSs in place of the damaged ones or as relay BS that relays the data to the nearest healthy node of the communication system.
This is an important vertical of UAV-assisted wireless networking, as in such scenarios, the transmitted message would be critical and can save the lives of people.

To this end, some key benefits of UAV-assisted wireless networking can be summarized as follows:
\begin{itemize}
    \item Ubiquitous connectivity can be provided to the users, as UAV-BSs are mobile and are capable of tracking the users.
    \item Pop-up scenarios are well managed in terms of  connectivity. 
    \item The business of mobile network operators becomes more sustainable as their capital expenditure (CAPEX) and operational expenditure (OPEX) are reduced because they do not need to deploy new fixed BSs since the same UAV-BSs can be reused in various occasions. 
    \item Emergency scenarios can be managed more efficiently, as the UAV-BSs provide a good amount of flexibility.
\end{itemize}

However, albeit its myriad of benefits, there are certain design challenges for UAV-assisted networking, including energy efficiency~(EE), trajectory planning, positioning, resource management, privacy, regulations, etc.~\cite{uav-sky, mozaffari2019tutorial}.
Of these, EE has a particular place, because UAV-BSs are mostly battery operated thereby their energy storage is limited.
Hence, the energy consumption becomes critical in order to: 1) prolong the flight time of UAV-BSs; 2) minimize CO$_2$ emission; 3) reduce CAPEX and OPEX of mobile network operators by making them require less UAV-BSs (since each can serve more) and bringing the energy costs down. 
Each of these items is quite important during the design process of the network.
Prolonging the flight time, for example, is not only instrumental in reducing the costs of mobile network operators but also effective in keeping the system fully operational.
In an extreme case where there is a very high demand, if a mobile network operator does not possesses sufficient number of UAV-BSs to replace one another once their batteries die, then there will be coverage holes and some users will be out of coverage, making the system partially down.
On the other hand, if the EE is in place, then the flight time of each UAV-BSs increases and that minimizes the time of without connectivity for some users.

This suggests that energy optimization should always be in the equation when it comes to UAV-assisted networking, meaning that even the earlier mentioned design challenges should also be coupled with EE through multi-objective optimization models.
Positioning, for instance, can be done not only to increase the number of connected users or to maximize the quality-of-service~(QoS) but also to minimize the energy consumption.
In other words, the optimum position of a UAV-BS can be determined in such a way that it can cover more users while consuming less power.
The main takeaway here is that the energy consumption of UAV-BSs requires special attention in order to make the concept more sustainable and feasible~\cite{uav-ml-survey,uav-survey-access}.

There is an avalanche of studies in the literature about the energy optimization for UAV-assisted wireless networking~\cite{gree-uav,Fotouhi2019}.
Because the nature of each problem is quite diverse, the studies in the literature employ various types of optimization techniques, including conventional (e.g., heuristics, game theory, etc.) and recent (e.g., machine learning~(ML)) ones.
Even though the conventional techniques are relatively stable and robust, we have witnessed a shift towards ML-based solutions, not only in energy optimization but also UAV-assisted networking in general~\cite{uav-ml-survey}. 
This mainly originated from the fact that ML techniques offer additional benefits, such as being adaptable and dynamic, which are vital features for wireless communication networks whose environments rapidly change.
On the other hand, error is an inevitable outcome of ML techniques~\cite{ml-survey}, and error rate and how much error can be tolerated are important design challenges.
Moreover, the availability, quality, and sufficiency of the data to be employed in training is a big issue for ML-based solutions~\cite{ml-chal-1,ml-chal-2}, overshadowing the above-mentioned advantages of ML. 
Therefore, it is obvious that each method---either conventional optimization theory or ML counterpart---has certain pros and cons, and it is not straightforward to point one method as superior to another; instead, they should be employed according to the conditions and requirements.

Apart from the optimization method being employed, there are also various and distinct perspectives to be focused on in terms of energy optimization for UAV-assisted wireless networking.
We can observe some well-studied concepts, including 1) positioning where the optimum three-dimensional positions of UAV-BSs are determined~\cite{Babu2021,Gao2019}; 2) trajectory planning where the paths of UAV-BSs are designed~\cite{Zeng2019,Tianyu2020}; 3) resource management~\cite{Meng2019,Meng2021}; 4) flight and transmission scheduling~\cite{Kang2020,Qi2020}.
In addition to these techniques, there is also a novel technique that recently attracted considerable attention, namely landing spot optimization, where UAV land on designated platforms to minimize propulsion energy due to flying or hovering, which is the most significant source of energy consumption~\cite{Petrov2020,BayerleinL2}.

In this survey, we first provide an overview of UAVs with top-level taxonomy in terms of their wing types~(i.e., fixed and rotary) followed by an in-depth analysis of the power supply and charging mechanisms of UAVs.
Considering the discussion on EE so far, the power supply and charging system, are at the heart of the UAV-assisted wireless networking.
Then, we thoroughly investigate the role of UAVs in wireless communication networks by presenting various use-cases, including backhauling, load balancing, capacity enhancement, etc.
With this discussion, we aim to reveal how broad the application range of UAVs in wireless communications is and that they will be at the core of 5G and 6G networks.
The types of UAV-BS deployments are also reviewed in this survey to distinguish standalone UAV deployment and UAV-BS deployment with fixed BSs.
In the former, there is no existing cellular network infrastructure available, while in the latter UAV-BSs assist the existing cellular network.
We then investigate the energy optimization phenomena in UAV-assisted wireless networking to showcase different types of energies to optimize, such as propulsion energy and communication energy.
This discussion is important to understand the main energy consumers in a UAV-BS, which specifies the energy optimization objectives accordingly.

As we mentioned earlier, there is a mountain of diverse optimization techniques available in the literature, and thus these techniques are surveyed here under the top-level taxonomy of conventional techniques (e.g., heuristics) and ML techniques.
The idea here is to capture general frameworks of those algorithms to better understand the state-of-the-art in the literature.
Then, energy optimization techniques---as the primary focus of this survey---are extensively surveyed under three-layer categorization: 1) deployment type; 2) energy optimization techniques (e.g., positioning, trajectory planning, etc.); and 3) the optimization method (e.g., conventional and ML).
Under the second layer (e.g., the optimization techniques), we included the above-mentioned landing spot optimization concept.
It is better to highlight this because it is quite novel and one of the most important contributions of this survey since, to the best of our knowledge, the landing spot concepts has never been covered in any survey work so far.
The three-layer categorization is performed in order to provide a better understanding with such a detailed mapping.

After that, we also discuss some enabling technologies for energy-efficient UAV-assisted wireless networking.
In that, we cover reconfigurable intelligent surfaces~(RIS), mobile edge and cloud computing, network slicing, and cooperative communications, which are quite timely and promising technologies.
This discussions aims to reveal such promising and emerging technologies and how they can be used in favour of EE.
Moreover, within this discussion, we also present the energy harvesting approach, which also has the potential of boosting the EE in UAV-assisted networking.
Even though it has been  quite extensive discussion so far with the inclusion of many perspectives, there is  still room for improvement in order to make the whole concept much more reliable, sustainable, and feasible.
We captured such improvement opportunities and various design challenges, such as security, data availability, regulations, etc., as well within the scope of this survey. 

\subsection{Related Works}\label{sec:related}
Since UAV-assisted wireless networking has attracted considerable attention due to the advantages mentioned above, in addition to the research works---which will be detailed later in this survey---, there are also numerous survey papers available in the literature.
In this section, we will present them and make a comprehensive comparison in terms of their content and focus.
It is worth noting that, among many, we carefully select the ones that are most relevant and recent in order to avoid diverting the scope of the discussion and keep this survey paper up to date with the latest literature.

The authors in~\cite{sr1} investigated how UAVs can be utilized in public safety communications, and they took the subject from the EE perspective.
After making an extensive classification of UAVs in terms of application, altitude, and network, they proposed a UAV network architecture for public safety. 
Then, the highlighted the critical need for EE in UAVs,
followed by various energy optimization techniques and optimization methodologies.
The authors did not go into much detail about the optimization algorithms and how they work, but they briefly discussed various methods.
A seminal paper in~\cite{mozaffari2019tutorial} gives a tutorial in UAV-assisted wireless networking.
In particular, UAVs were first classified in terms of their wing types (e.g., fixed and rotary) and altitude (e.g., low and high).
Then, UAV assistance in wireless networks was thoroughly discussed by considering their roles in the next generations of wireless networks under different scenarios, including UAVs as flying BSs and UAVs as flying user equipment.

Artificial intelligence~(AI) integration to UAV-assisted wireless networking was analyzed in~\cite{sr3} with brief tutorials for different ML algorithms from different categories, such as supervised, unsupervised, and reinforcement learning (RL).
Intelligent reflective surfaces~(IRS) combined with UAVs were discussed, as well as how RL can be utilized for optimization purposes.
Federated learning (FL), its advantages, and its applications to UAV-assisted wireless communications were also included with a detailed discussion.
A similar survey paper in~\cite{uav-ml-survey} also revealed how ML would help make UAV-assisted communications more effective.
Starting with the introduction of the characteristics of UAVs, AI, and ML, the authors grouped the role of ML for UAV-assisted networks into four different categories: namely, physical layer aspects, resource management, positioning, and security.
Various subjects, such as interference management, data caching, jamming, and mobility, were discussed separately under each category with an ML focus.

Albeit not discussed from a wireless communication perspective, the survey paper in~\cite{sr5} focuses on deep learning (DL), a sophisticated version of ML, applications for UAV systems.
The authors extensively reviewed DL algorithms and their uses in UAV systems for different applications, including motion control, situational awareness, and path planning for search and rescue missions.
The authors concluded their work with the challenges in both DL and UAV autonomy.
The short review paper in~\cite{sr6} tried to reiterate the importance of self-organization\footnote{For a detailed discussion on self-organization in cellular networks, the readers are referred to~\cite{son-survey}.} in UAV-assisted connectivity.
The scalability issue in UAV-assisted networking was raised, and the role of distributed algorithms in mitigating such issue was also discussed.

The survey in~\cite{gree-uav}---being the closest one to our survey---, mainly investigated the concept of green networking for UAV-assisted 6G networks.
Different approaches in green UAV communications, including energy saving, energy harvesting, and RIS-assistance, were presented, followed by the enablers of green UAV communications.
The functionality of UAV-assisted green 6G networks were also analyzed, and new research directions in order to make the UAV-assisted 6G networks more feasible were identified.
The authors in~\cite{Fotouhi2019} provides a detailed survey in terms of UAV cellular communications.
Unlike the survey papers mentioned above, the standardization issue involving the study item phase by the third generation partnership project~(3GPP) was also covered.
The authors also included the 3GPP work item phase along with some non-3GPP standardization initiatives, such as The International Telecommunication Union Telecommunication Standardization Sector~(ITU-T), The European Telecommunications Standards Institute~(ETSI), and The Institute of Electrical and Electronics Engineers~(IEEE).
Such in-depth standardization discussion makes the survey in~\cite{Fotouhi2019} unique and quite informative in that sense.

\subsection{Motivation, Contributions, and Organization of the Survey}
There is a myriad of survey papers on UAVs and their roles in different domains, including agriculture, surveillance, search and rescue, etc.
As discussed so far in this survey paper, their roles in wireless communications are also phenomenal due the additional degrees of freedom that they provide.
In this regard, the number of studies investigating the assistance of UAVs in wireless communications is huge, and there is also a considerable amount of survey papers that bring the studies in particular perspectives together.

Since  energy optimization lies at the heart of UAV-assisted wireless communication networking, we intend to produce a dedicated survey paper on this topic in order to compile the research efforts in energy optimization, with three main objectives:
\begin{itemize}
    \item to highlight the cruciality of energy optimization in UAV-assisted wireless networking.
    \item to reveal the state-of-the-art in order to understand where we currently stand.
    \item to identify the gaps in the literature, which further research should focus on.  
\end{itemize}

Furthermore, given that ML has proven to be an efficient and effective tool in wireless communications networking~\cite{unsup}, it has been widely used in UAV-assisted networking as well~\cite{uav-ml-survey,sr5, abubakar2020role}.
Therefore, the application of ML in energy efficient UAV-assisted networking is also included in this survey in order to capture the current research trend in the literature.
In addition, conventional optimization algorithms, such as genetic algorithm, game theory, and particle swarm optimization, are covered in this survey since they also have a large application area in the case of energy efficient UAV-assisted networking.
The primary intention of including both the conventional and ML methods is to provide a holistic approach that does not prioritize one over another, given that the corresponding literature itself have studied all types of optimization methods.
Therefore, with this approach, the literature is better represented.

To this end, although there are many survey papers on UAVs in the literature---some are detailed in Section~\ref{sec:related}---, to the best of our knowledge, there is no survey paper that precisely investigates the energy optimization methods for UAV-assisted wireless communication networking by combining the conventional and ML methods.
To be more specific, even though almost all the surveys include a small or large amount of content on energy consumption in UAV-assisted wireless networking~\cite{sr1, mozaffari2019tutorial, sr3, gree-uav}, none of them has the energy optimization methodology as the sole focus.
The work in~\cite{gree-uav} is the closest to our survey as its primary focus is also green UAV-assisted networking; however, our survey has a number of differences:
\begin{itemize}
    \item The focus of our survey is not only on energy optimization in general but also the optimization methods employed in energy efficient UAV-assisted wireless communication networking.
    \item We categorized the methods according to their type~(i.e., conventional and ML) and investigated each energy optimization method accordingly.
    \item For the sake of completeness. A brief tutorial about the optimization methods is included in our survey.
    \item We also included the novel landing spot approach, which has gained momentum in the research community.
\end{itemize}

The summary of contributions and the paper organization is as follows:
\begin{itemize}
    \item Section~\ref{sec:types} discusses different types of UAVs in order to reveal their characteristics and capabilities, which is quite important in selecting the UAV for a particular application.
    \item The power supply and charging mechanisms of UAVs are extensively covered in Section~\ref{sec:power_supply}.
    This is particularly important because the optimization can be performed according to the power supply (e.g., battery, grid, fuel, renewable, hybrid), and various charging/recharging mechanisms (e.g., battery swapping, refuelling, wireless power transfer, etc.) can be placed into the optimization model.
    \item The role of UAV-BSs in wireless communication networks are investigated in Section~\ref{Sec:Four}.
    Since this survey is oriented around wireless communication networks, we include a thorough discussion on how UAVs can help and what their primary use-cases are.
    Such discussion also reiterates the reasoning behind using UAVs in today's and future wireless communication networks and somehow uncovers the importance of the efforts trying to make the whole concept feasible.
    \item Section~\ref{sec:deployment} presents different types of UAV deployments in order to explain the difference between standalone UAV deployments and UAV-assisted cellular networking because, according to this, the energy optimization model changes significantly.
    \item The energy optimization in UAV-assisted wireless networking is covered in Section~\ref{sec:EnergyOpt}, wherein the energy optimization is categorized according to the optimization objective (i.e., propulsion energy, communication energy, and joint optimization of propulsion and communication energies).
    \item For the sake of completeness of this survey paper, the overview of both conventional and ML algorithms employed in energy optimization of UAV-assisted wireless networks is given in Section~\ref{sec:algorithms}.
    \item The energy optimization techniques, as the core part of this survey, are thoroughly discussed in Section~\ref{Sec:eight}. 
    Various techniques are introduced by presenting the related literature, which is categorized in terms of the type of UAV deployment and optimization method employed (e.g., conventional and ML).
    With this section, the state-of-the-art is demonstrated, and a recently proliferating concept---called landing spot optimization---is also included to capture the energy optimization in UAV-assisted wireless networks holistically.
    \item To understand what and how other technologies can boost EE in UAV-assisted wireless networks, Section~\ref{sec:enablers} mainly introduces the enabling technologies.
    In this section, a novel technology, called RIS is included as one of the enablers, since RIS has recently gained a significant amount of interest in the research community.
    \item Section~\ref{sec:challenges} identifies the primary challenges and possible future research directions in order to fill in the gaps in the literature that would enable the overall UAV-assisted wireless communications concept to be more feasible.
    Lastly, Section~\ref{sec:conclusion} concludes the survey with final remarks.
\end{itemize}

\begin{table}
\centering
\caption{List of Acronyms}
\label{tab:my-table}
\begin{tabular}{|l|l|}
\hline
Acronym & Full Meaning                 \\ \hline
3D      & Three Dimensional       \\ \hline
5G      & Fifth Generation        \\ \hline
6G      & Sixth Generation        \\ \hline
ADMM    & Alternating Direction Method of Multipliers                       \\ \hline
AI      & Artificial Intelligence \\ \hline
ANN        & Artificial Neural Networks                        \\ \hline
AP      & Access Point            \\ \hline
ARIMA   & Auto-Regressive Integrated Moving Average                        \\ \hline
AtG     & Air-to-Ground           \\ \hline
BCA       & Block Coordinate Ascent                        \\ \hline
BCA       & Block Coordinate Descent                        \\ \hline
BS      &  Base Station           \\ \hline
CA      & Conventional Approaches                        \\ \hline
CAPEX   & Capital Expenditure     \\ \hline
CCP       & Concave Convex Procedure                        \\ \hline
CNN        & Convolution Neural Networks                        \\ \hline
C-RAN    & Centralized Radio Access Network                        \\ \hline
CS       & Cucker-Smale                        \\ \hline
D2D     & Device-to-Device        \\ \hline
DBS     & Data Base Station                      \\ \hline
DDPG    & Deep Deterimistic Policy Gradient                      \\ \hline
DRL     & Deep Reinforcement Learning                      \\ \hline
EE     & Energy Efficiency                       \\ \hline
GMM     & Gaussian Mixture Model                       \\ \hline
GPS       & Global Positioning System                        \\ \hline
HIL     & Hardware in Loop                        \\ \hline
FL      & Federated Learning                         \\ \hline
FSO     & Free Space Optics                       \\ \hline
kNN     & k-Nearest Neighbour     \\ \hline
IoT     & Internet of Things                       \\ \hline
LiPo    & Lithium Polymer           \\ \hline
LoS     & Line of Sight                         \\ \hline
LSTM    & Long and Short Term Memory                       \\ \hline
MBS     & Macro Base Station   \\ \hline
MDP     & Markov Decision Process                        \\ \hline
MEC     & Mobile Edge Computing                        \\ \hline
MIMO    & Multiple Input Multiple Output                       \\ \hline
ML      & Machine Learning                        \\ \hline
mmWave  & Millimeter Wave                        \\ \hline
MNIST   &  Modified National Institute of Standards and Technology                       \\ \hline
NFV     & Network Function Virtualization                        \\ \hline
NLP     & Natural Language Processing                      \\ \hline
NR      & New Radio                 \\ \hline
NOMA    & Non-Orthogonal Multiple Access \\ \hline
OEM     & Original Equipment Manufacturer                        \\ \hline
OPEX    & Operating Expense         \\ \hline
PSO     & Particle Swarm Optimization                       \\ \hline
PV      & Photo voltaic                        \\ \hline
QoS     & Quality of Service        \\ \hline
RF       &  Radio Frequency                       \\ \hline
RIS     & Re-configurable Intelligent Surfaces                       \\ \hline
RL     &  Reinforcement Learning                     \\ \hline
RNN     & Recurrent Neural Networks                      \\ \hline
SARSA   & State Action State Action Reward                      \\ \hline
SCA     & Successive Convex Optimization                        \\ \hline
SMPS    & Solar Power Management System                        \\ \hline
SVM     & Support Vector Machine \\ \hline
SVR     & Support Vector Regression                     \\ \hline
TBS     & Terrestrial Base Station                       \\ \hline
TDMA    &  Time Division Multiple Access                       \\ \hline
THz     & TeraHertz             \\ \hline
T-UAV   & Tethered Unmanned Aerial Vehicle                        \\ \hline
UAV     & Unmanned Aerial Vehicle                        \\ \hline
URLLC   & Ultra-Reliable Low Latency Communication                      \\ \hline
WEM     & Weighted Expectation Maximization                      \\ \hline
WiFi       &  Wireless Fidelity                       \\ \hline
WPCN    & Wireless Powered Communication Networks                       \\ \hline     
WPT     & Wireless Power Transfer                        \\ \hline
XGBOOST & Extreme Gradient Boosting                       \\ \hline
\end{tabular}
\end{table}

\section{Types of UAVs}\label{sec:types}
UAVs, also known as drones, are of two main categories: fixed-wing and rotary-wing.  However, with the advancement in UAV technology and the wide range of applications of UAVs, rotary-, and fixed-wing UAVs can be combined to form a hybrid design~\cite{Fotouhi2019}.  
\par
Rotary-wing UAVs are designed to perform vertical take-offs and landings. One of the main design features of rotary-wing UAVs is that they can hover on a fixed and specified location, making them perfect candidates to perform tasks like continuous cellular coverage and sensing~\cite{Fotouhi2019}.  
However, rotary-wing UAVs consume more energy since they operate at a low altitude with little mobility, and their constant flight against gravity results in more power consumption~\cite{Fotouhi2019}. 

Fixed-wing UAVs are another type of UAV that can glide through the air and operate at higher altitudes, making them more energy efficient and capable of carrying heavier payloads. Moreover, fixed-wing UAVs, such as tiny planes, have heavier weights, faster speeds, and must move forward to stay airborne~\cite{mozaffari2019tutorial}. However, fixed-wing UAVs require a runway for landing and take-off and are more expensive than rotary-wing UAVs~\cite{nonami2010autonomous}.

The limitations of both rotary-, and fixed-wing UAVs led to the emergence of a new type of UAV in terms of shape and aerodynamics, called hybrid UAVs~\cite{saeed2015review}.  
The fundamental design strategy behind the hybrid ones is to combine the design features of both rotary-, and fixed-wing UAVs. 
Hybrid UAVs employ different features of both rotary-, and fixed-wings for various maneuvers and flights dynamics. These UAVs can perform vertical takeoff and landing (VTOL) in copter mode and shift to high-speed forward flight in aeroplane mode~\cite{ducard2021review}. For example, a parrot swing UAV~\cite{Fotouhi2019}, which is an improved version of the traditional four-arm quadcopter~(a rotary-wing UAV), has been equipped with some fixed-wing UAV features. As such, it can take off vertically quickly, hover, and fly horizontally at super-sonic speed.

\section{UAV power supply and charging mechanisms}\label{sec:power_supply}
\begin{figure*}[t!]
	\centering
	\includegraphics[width=0.95\textwidth]{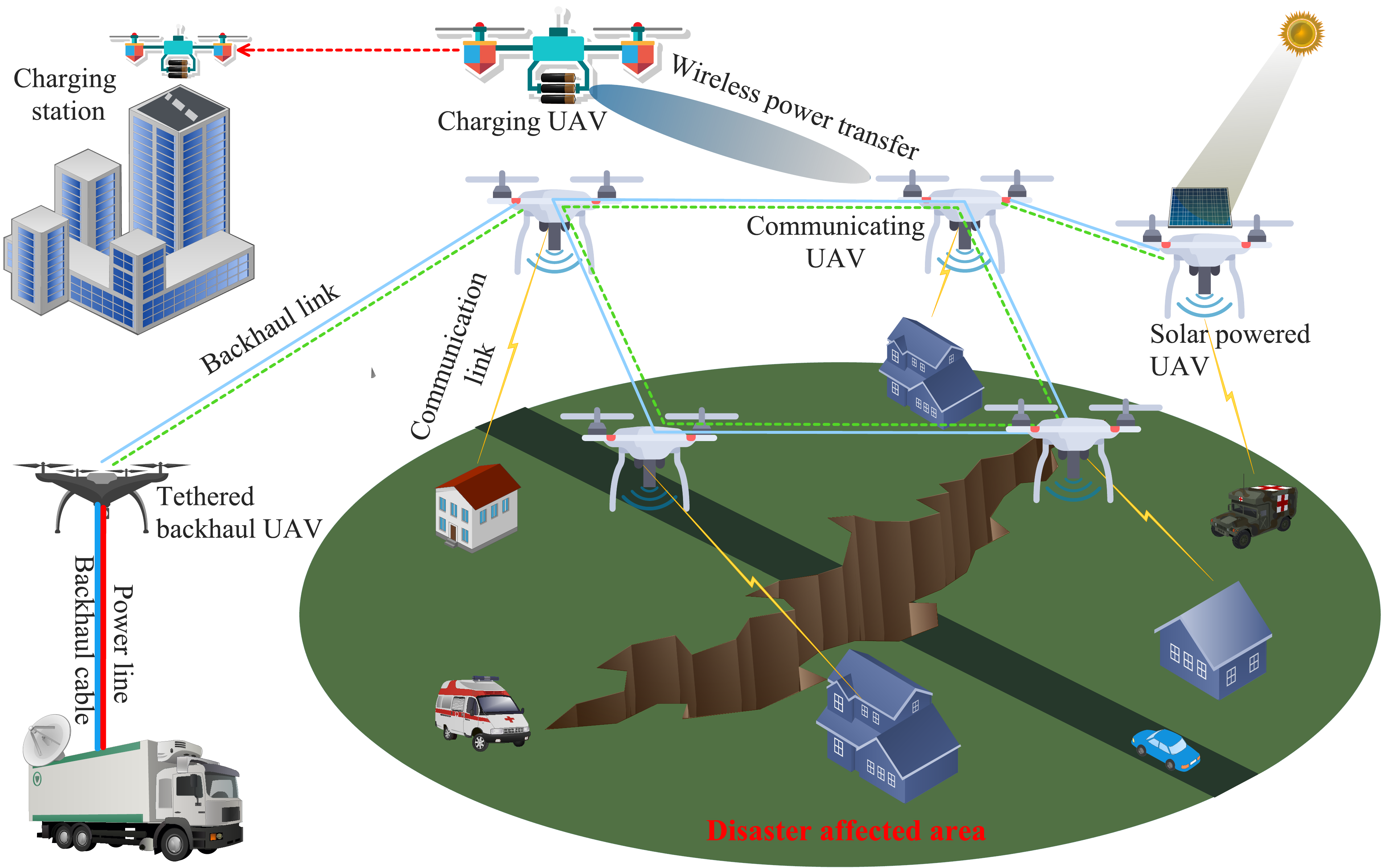}
	\caption{Various UAV power supply and charging mechanisms including wireless power transfer via a recharging UAV, a tethered UAV, a solar powered UAV and battery swapping at a charging station.}
	\label{fig:drone_power}
\end{figure*}

In recent years, there has been an increase in the application of UAVs in both commercial and military domains, due to their easy adaptability, flexibility of deployment, and  cost-effectiveness~\cite{mozaffari2019tutorial}. 
However, to fully exploit the capabilities of UAVs in different application domains, there is a need to consider the challenges and limitations of UAVs. One of the most important limitations in UAVs is power and energy consumption~\cite{Fotouhi2019}. Commercial UAVs are usually powered with rechargeable batteries for operations, while large UAVs, such as military UAVs use non-renewable resources, such as fuel and gas, to provide more energy to the UAV for longer flight time. 
In recent years, new and alternative methods for UAV power supply, including solar energy, wireless charging, laser beam charging systems, etc., have been developed and tested~\cite{boukoberine2019critical}. In the following paragraphs, we discuss different methods of power supply and charging mechanisms for UAVs while Fig.~\ref{fig:drone_power} illustrates various UAV power supply and charging mechanisms.

\subsection{Battery Powered UAVs} The battery power supply is one of the main power supply techniques used to meet the energy requirements of UAVs. Batteries are mostly preferred in relatively smaller UAVs because classical batteries, such as lithium polymer (LiPo), can power a UAV for a maximum of 90 minutes\cite{khofiyah2018goldsmith, verstraete2012design}. 
This severely limits the commercial and industrial applications of battery-operated UAVs. 
Albeit significant advancements in battery technology, the limitations associated with the use of batteries to power UAVs are still far from over. Since batteries are the most predominant source of energy for miniature UAVs, and the energy stored in the battery is limited and can be easily depleted during the UAV operation, to increase the flight time and operation of UAVs, batteries must be frequently charged in order to replenish the depleted energy stored in them. To address such limitations, several battery charging techniques have been developed. Hence, in the following, we consider different battery charging mechanisms devised to recharge the UAV battery and ensure the longevity of UAV operation. 

\subsubsection{Battery swapping} 
The swapping process consists of recharging or replacing the UAV batteries and can be done conventionally or via hot swapping. In the conventional swapping method, the UAV whose battery is depleted has to leave its service location to the charging station and be replaced by an already charged UAV. 
The challenge with this method is that it requires several backup UAVs to be able to replace the UAVs whose batteries are depleted; as pointed out in~\cite{galkin2019uavs}, up to two standby UAVs may be needed per UAV to ensure continuous coverage for a commercial UAV that has a charging power of 180 V. However, the exact number of backup UAVs would depend on the UAV downtime, which comprises both the time taken for the UAV to fly to and from the charging station as well as that taken to recharge the battery to full capacity~\cite{boukoberine2019critical}. 

In the hot swapping approach, the UAVs do not have to be powered off or remain in the charging station until the battery is recharged, instead, as soon as they reach the charging station, new batteries are quickly inserted into them so that they can return to their operating station immediately. The limitation of hot swapping is that it requires human involvement to replace depleted batteries with new ones. To address this challenge, automated battery swapping mechanisms have been developed whereby a robotic actuator can be used to remove and replace UAV batteries~\cite{Danny2015}. 
For swapping to be implemented effectively, a battery recharging station is required to recharge or replace depleted batteries. 
This recharging  station can be located on cellular towers, rooftops of buildings, and specialized standalone pylons. There is also the need for multiple UAVs to ensure continuous service provision. In addition, a management system must be put in place to coordinate the battery recharging and replacement cycle of the swarm of UAVs~\cite{boukoberine2019power}. 

\subsubsection{Laser beam charging} Another technique for battery recharge is laser beam in-flight recharging \cite{boukoberine2019power}. This technique involves recharging the UAV battery during the flight, without making it land for the battery recharging or replacement, as is the case in battery swapping technique. 
To recharge a UAV's battery using the laser beam technique, an external energy source feeds the laser beam at a particular frequency and wavelength. The laser is directed towards a photo-voltaic~(PV) cell on the UAV that converts the laser beam into electrical energy required to recharge the UAV's batteries. To ensure maximum output from laser beam charging technique, a power tracking device is used to identify the maximum energy point in the laser beam. In order to maintain power transfer from the laser transmitter, the UAV must operate at low heights and occasionally in a restricted area while using the laser-beam inflight recharging strategy \cite{boukoberine2019power}.
Furthermore, because each UAV requires its own dedicated laser source, the number of UAVs must be reduced in order to maintain a fair operational cost \cite{galkin2019uavs}
The laser beam is an effective method for charging UAVs, and it can be applied to both rotary-, and fixed-wing UAVs.  However, the use of laser beam recharging methods restricts the operation of the UAV to a limited area in order to ensure that the recharging link is active. 

\subsubsection{Wireless power transfer/wireless charging}
Wireless power transfer (WPT) is another technique for countering and overcoming the limitations of conventional battery power supply mechanisms to UAVs. The concept of wireless charging was introduced by Nikola Tesla~\cite{simic2015investigation}.  To perform the wireless charging of electronic devices, a transmission pad is used to charge the device using the resonant inductive coupling \cite{simic2015investigation}. A typical WPT consists of two main components: namely, transmitting device and a receiving device. A transmission device is used for the conversion of the source energy into “time-varying electromagnetic fields” which is then transmitted using a transfer media, which converts the time-varying electromagnetic arrays into electrical power that is, in turn, used to power the UAV. 

Currently, the WPT technique is in its early-stages and there are still several issues and limitations that needs to be addressed. 
For instance, due to the significant propagation loss of RF signals over long distances, the performance of WPT systems with a wide coverage range is essentially hampered by their low end-to-end power transmission efficiency. As a result, fixed-location energy transmitters~(ETs) must be distributed in an ultra-dense way in order to provide pervasive wireless energy access for enormous low power  energy receivers (ERs) deployed over a wide area. 

Using WPT, some UAVs can serve as recharging UAVs to ensure that batteries of other UAVs---that are serving user requests---are not depleted by flying to their respective locations to recharge their batteries while they are in operation. This will prevent service interruptions that might occur if each UAV has to fly to a charging station located at a distance from the UAV operating region to replenish its depleted battery. In such an arrangement, there is normally a transmitting UAV~(tUAV) whose responsibility is to transmit RF signals to the serving UAVs~(sUAVs), which have been equipped with sensors to convert the received RF signals to electrical energy needed to power the UAVs~\cite{Jahan2019}. 
 
\subsection{Grid Powered UAVs: Tethering}
Another technique to power UAVs is called tethering, which allows the connection to be established between the UAV and power supply on the ground using a fiber optic cables\cite{boukoberine2019power}. Fiber optic connections allow several kilowatts of power to be transferred using high-intensity light. This technique provides more autonomy to UAVs and allows secure and quick data transfer in real-time. 
However, the tethered UAVs~(TUAV) are only allowed to hover within a certain hovering region since the tether has a maximum length and the launching point of the TUAV is typically located on a rooftop. 
A study in \cite{kishk20203} investigated the best location for tethered UAVs (TUAVs) to reduce the average path-loss between the TUAV and a terrestrial receiver. They calculated the upper and lower bounds for appropriate tether length and inclination angles in order to reduce average path-loss. 
Asides the maximum tether length, the heights of the buildings surrounding the TUAV rooftop determine the hovering region, which necessitates that the tether's inclination angle does not fall below a certain minimum value to avoid tangling and to ensure safety. It is worth noting that if the tethered cable delivers constant power, the UAV flight is expected to last a few days or possibly a few months~\cite{gu2016novel}.

Tethered UAVs have been considered for various wireless network applications in the literature. The work in \cite{bushnaq2020optimal} gives a comparative performance analysis of untethered-UAV~(U-UAV) and T-UAV-assisted cellular traffic offloading in a traffic-heavy location. They employed stochastic geometry methods to calculate joint distance distributions between users, the terrestrial BS~(TBS) and UAV. A user association strategy is presented, and the relevant association areas are analytically discovered to optimize the end-to-end signal-to-noise ratio~(SNR). The total coverage probability of U-UAV/T-UAV-assisted systems is then calculated for specified TBS and U-UAV/T-UAV locations. The authors in~\cite{matracia2021topological} introduced a novel UAV-based post-disaster communication system where U-UAVs are employed to offer cellular service in  disaster-affected areas, while T-UAVs provide backhaul for the U-UAVs.

\subsection{Fuel Cell Powered UAVs}
Fuel cell is also used to power UAVs; the use of fuel cell is more effective and efficient as compared to the battery power UAVs, such that a fuel cell power increases the flight time of UAVs six times as compared to battery \cite{hwang2013lifecycle}. However, the fuel cells have their own limitations; for example, the fuel cells have lower energy density and require special consideration for fuel tanks~\cite{pan2019recent}. Hydrogen cannot be stored at high pressure and lower temperatures. To overcome fuel cell density issues, compressed hydrogen gas, liquid hydrogen or chemical hydrogen can be used in fuel powered UAVs \cite{kendall20124}. Fuel cells are much lower than lithium batteries in terms of power storage and efficiency, according to \cite{pan2019recent}, 
fuel cell can achieve a maximum level of efficiency up to 60\% as compared to the lithium batteries that achieve an efficiency level of 90\% in terms of power storage \cite{boukoberine2019critical}. The limited power storage capacity in fuel cells is due to the required auxiliary subsystem needed for fuel cell stacking operations.
To tackle this issue, the authors in \cite{rhoads2010design} used compressed hydrogen to power the UAV, resulting in a total flight time of about 24 hours.

The generation of chemical hydrogen requires a specialized infrastructure and equipment, making the power supply system very complex and heavy. Moreover, the process of hydrogen extraction usually takes more time, leading to increase in response time of the UAV to load changes, which creates load balancing problems. 

\subsection{Renewable Energy Powered UAVs}
Various types of renewable energy (e.g., wind, solar, etc.,) can be used to power UAVs in order to increase their travel time and power efficiency. To power UAVs with wind energy, gust soaring can be used by adjusting the trajectory of the UAVs, which enables the UAV to extract the energy from the wind by converting the potential energy of the wind into kinetic energy \cite{richardson2015upwind}.
One of the main limitations is the dependence on the environmental condition and airflow~\cite{bonnin2015energy}. 
Another prominent technique to power UAVs is to use solar PV cells that can be mounted on the wings of the UAVs in order to recharge them via the irradiation from the Sun. One of the main drawbacks of this type of power supply is that it limits the flight of UAVs in rain and during nighttime, when there is less or no irradiation from the Sun. Hence, the unavailability of the sun in the nighttime and rain entails that the UAVs needs to be powered with another form of power supply~\cite{oettershagen2015solar}.

\subsection{ Hybrid Powered UAVs}
Hybrid power supply methods can be used to power UAVs, and they combine battery, fuel cells, and renewable energy sources to provide a blend of power supply \cite{donateo2017new}.

\subsubsection{Fuel cell-Battery}
There are some limitations associated with the use of fuel cell, and battery-powered UAVs. 
On one hand, the process of electricity generation from fuel cells takes a long time as several components such compressors for air supply, pumps and valves are involved, which leads to increase in response time of the UAV to load changes. On the other hand, battery-powered UAVs have limited flight time and would require frequent recharging, which can negatively affect the effectiveness of their service. Therefore, hybridization of fuel cells and batteries can be used to minimize the delays associated with electricity generation as well as prolong the service time and effectiveness of UAVs \cite{belmonte2018fuel}.

Hybrid power supply resources can be used interchangeably to power UAVs. For instance, battery power can be utilized by UAVs during take-off and ascending process because batteries have more density and power storage as compared to fuel cells, and UAVs require more power when taking off and ascending to higher altitudes \cite{sun2019optimal}. Then, fuel cell can be used in the flight time and descending process. Furthermore, the fuel cell can be used to charge the batteries. In \cite{verstraete2012hardware}, the authors used a technique called hardware in the loop (HIL) in the hybrid UAVs to investigate the power consumption of fuel cells and battery power sources. Several simulations were conducted to determine the level of flight endurance that can be achieved. Another study was conducted by~\cite{gong2014role} using fuel cells and battery power supply to assess the battery power contribution under different conditions and scenarios.

\subsubsection{Solar Cells plus Battery}
Another method for extending the mission length of UAVs is the installation of PV panels to operate alongside the existing batteries. When solar irradiation is available, PV cells are often employed to power a UAV or refill its battery. The battery, in turn, is utilized for functioning at night or during hours when solar radiation is limited~\cite{lu2018wireless}. The use of solar power as an energy source allows small size UAVs to carry larger payloads and can increase flight periods to more than 24 hours, allowing for multi-day flying \cite{morton2015solar}. 
The authors in~\cite{shiau2009design} studied the design and validation of a solar power management system (SPMS) for their solar-cell and battery-powered experimental UAV. Their results reveal that when the angle of incidence of sunlight varies from 0 to 45 degrees, the power consumed from the solar cells fluctuates depending on the load situations and may be reduced by up to 30\%. This means that changes in aircraft attitude will have a direct impact on the power generated by the solar system.  

\section{The role of UAV-Base Station in Wireless communications}\label{Sec:Four}
\begin{figure*}[t!]
	\centering
	\includegraphics[width=0.99\textwidth]{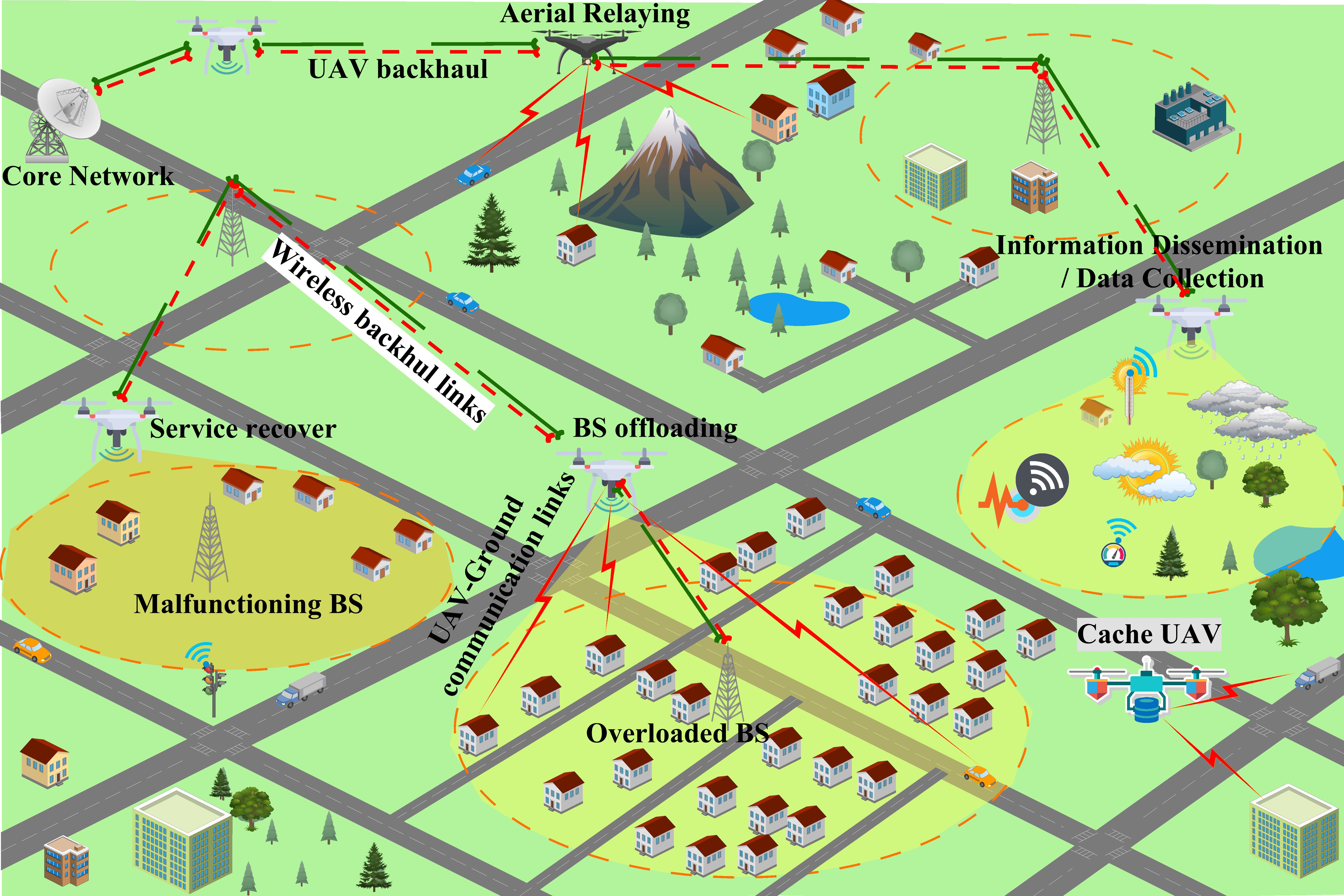}
	\caption{Different UAV use cases involving the UAV serving as a \textbf{wireless backhaul} to connect users located in remote areas to the core network, as a \textbf{emergency network.} for restoring service to an area covered by a malfunctioning BS, as a \textbf{aerial relay} for providing coverage extension to users separated by obstacles such hills and mountains, \textbf{data offloader} for offloading user traffic from overloaded terrestrial BSs to ensure load balancing and throughput enhancement, as \textbf{aerial cache} for serving popular content to ground users to reduce backhaul congestion and achieve EE and as a \textbf{data harvester} for collecting data from ground sensor nodes.}
	\label{fig:drone_use_case}
\end{figure*}

UAVs have been employed for various operations in both military and civilian domains, including object detection, location tracking, goods delivery, disaster monitoring, information dissemination, etc~\cite{mozaffari2019tutorial}. Recently, they have also found several applications in wireless communications because of their flexibility, adaptability and easy deployment~\cite{Fotouhi2019}. Fig.~\ref{fig:drone_use_case} shows various use cases of UAV-BSs in wireless communications. In the following paragraphs, we briefly discuss some of these applications.

\subsection{Emergency Services~(Pop-up Networks)}
UAVs are becoming one of the most active areas of research and industrial development due to their promising applications in different domains. One of the application areas for UAVs' usage is emergency response~\cite{Sambo2019}. During large scale  natural disasters like floods, earthquakes, or major fire outbreaks, hurricanes, etc., several properties and infrastructures are normally destroyed, including cellular infrastructures, thereby leading to loss of communication service in that area~\cite{ jin2020research}. In such situations, UAVs can be deployed to such areas to replace the malfunctioning BSs and provide emergency communication service. This would help to improve the QoS of the users whose network service was affected by the natural disaster. UAV parameters such as trajectory, altitude, etc., can be optimized based on the user traffic demand and distribution to maximize their throughput and coverage. 
In this regard, the work in~\cite{paulo-uav} considered the deployment of UAVs as emergency pop-up networks to restore communication services to users in an area where due to natural disasters, many cellular network infrastructures have been destroyed. Then, they applied RL to optimize the trajectories of the UAVs in order to maximize the coverage and throughput of ground users.

In addition, the UAV can serve as a relay to connect an isolated group of users that are separated from each other or provide backhaul services to other existing wireless networks such as device-to-device~(D2D) communications, etc~\cite{Deepak2019}.
They can also be used to acquire and process real-time data and information using remote sensing, remote control technology, and other communication technologies (e.g., IoT). This real-time information can be used for decision-making by emergency response teams and decision-makers because the quick response time and accurate decision making are important factors in carrying out effective rescue mission during emergencies~\cite{aasen2018quantitative}.

\subsection{Data Harvesting from IoT Devices}
Due to the promising solutions that UAVs provide in different application areas, UAVs has attracted research attention in IoT applications in order facilitate data harvesting and relaying~\cite{singh2021proficient}. This is very important for cases whereby the IoT devices are deployed in areas where there are no wireless network facilities, such as in rural farmlands or offshore locations, to help transmit the data to the decision-making centres. In such situations, UAVs can be deployed to that location to assist in data collection~\cite{tsouros2019review, maddikunta2021unmanned}. Research works on UAV-enabled data harvesting mainly focus on UAV path planning and throughput maximization in order to ensure that the UAV collect sufficient data and returns to data centre before its battery depletes~\cite{Tianxin, Bayerlein2020H, liu2020uav}.

Following this research direction, a deep RL  framework for path planning in a multi-UAV based data harvesting system that can adapt to dynamically changing network parameters, such as varying number of UAVs and IoT devices, different flight schedules, and amount of data to be harvested, was proposed in~\cite{Bayerlein2020H}. The path planning problem was modelled as a cooperative team of UAVs saddled with the responsibility of maximizing data collection from several IoT devices distributed in an area. Then a multi-agent $Q$-learning algorithm was developed to determine the optimal trajectories that would maximize data collection from the distributed IoT devices.
In~\cite{Tianxin}, the authors considered the joint optimization of both the UAV trajectory and power allocation to the ground nodes in order to maximize the throughput and coverage probability in a UAV-enabled data harvesting system with distributed beamforming. Heuristic algorithms based on convex optimization and approximations were developed to find the optimal trajectory and power allocation strategy that would maximize both performance metrics.

\subsection{Content Caching and Computation Offloading}
Content caching is an important process in modern-day wireless communication networks as users move from one location to another. It involves storing important information like username, location and popularly requested content at multiple BSs in order to provide seamless communication and minimize the latency involved in information retrieval and transmission to the user~\cite{Dong2016Liu}. However, most of the caches are usually installed in a fixed location and would not be suitable for highly mobile users in vehicles and high-speed trains as the requested content would have to be stored in all BSs along the user path. In such situations, dynamic caching can be achieved by mounting the caches on UAVs, in what is known as UAV enabled caching~\cite{DAI2020}, and making the UAVs follow such highly mobile devices to provide the requested services. 
In this regard, the authors in~\cite{Huaqing2020wu} proposed a learning-based joint caching and trajectory optimization scheme in vehicular networks in order to enhance the throughput of the network.

Mobile edge computing~(MEC) was developed to assist user devices in offloading and processing computation-intensive tasks that are beyond their battery capacity~\cite{Pavel2017}. However, these MEC servers are usually deployed in a fixed location within the network and may not always be accessible. To solve this problem, UAV enabled MEC has been proposed in~\cite{Cui2019Yan} where UAVs equipped with edge servers can assist in computation offloading from ground user devices in order to minimize their energy consumption and maximize their battery lifetime as well as respond to real-time computationally intensive data processing demands. Regarding this, the authors in~\cite{Hui2020} proposed a UAV enabled computation offloading framework based on deep RL to minimize the network latency, energy, and bandwidth cost.

\subsection{Load Balancing} Due to the movement of mobile users from one place to another as well as variation in user traffic demands, the traffic loads of BSs vary both temporally and spatially. This makes some BSs to be lightly loaded while others are heavily or overloaded, thereby leading to service denial or poor QoS from the overloaded BSs~\cite{Anuradha2015}.  Although small cells helped to reduce the level of traffic imbalance in cellular networks, due to the static nature of their deployment, they are not able to respond to sudden changes in traffic demands that may be occasioned by large events taking place in an area for a short time or other impromptu surge in network traffic that requires quick intervention to reduce the pressure on the available BSs. In such cases, UAV-BSs can be of great help because they can be quickly deployed to respond to such traffic demands and ensure that load is properly distributed among the various BS in order to avoid network congestion~\cite{Irem2016}. 

In this regard, the authors in~\cite{JunShi2019} proposed a learning-based framework for UAV deployment to crowded regions of the network during periods of peak traffic to assist in load balancing, thereby preventing degradation in the QoS of users. In the proposed framework, both auto-regressive integrated moving average~(ARIMA) and extreme gradient boosting~(XGBOOST) ML algorithms were employed to forecast future high traffic intensity regions based on historical data for proactive UAV deployment. The work in~\cite{Qiang2019} considered UAV deployment for load balancing to reduce communication latency between IoT devices and macro BSs~(MBSs) during periods of high traffic load on the MBS. Heuristic algorithms were developed to determine the optimal location of the UAVs, and association strategy for the IoT devices.

\subsection{Coverage Extension/Relaying} The fast growth of mobile devices, such as smartphones, tablets, and wearables, has increased the demand for high-speed wireless access. As a result, the capacity and coverage of existing wireless cellular networks have been overstretched, thereby leading to the development of new wireless technologies to address this issue \cite{mozaffari2019tutorial}. Because of their vast range of use-cases, UAVs have sparked many attention \cite{hayat2016survey}.
The air-to-ground (AtG) dominant line-of-sight (LoS) link enhances the performance of UAV-enabled wireless communications \cite{al2014modeling}, thereby resulting in reduced propagation loss and better link QoS. As a result, UAVs functioning as aerial BSs or relays are commonly employed to increase network capacity  or achieve more flexible coverage \cite{jaziri2016congestion, zhan2011wireless, zeng2016throughput}.

The authors in~\cite{li2019throughput} investigated  mobile relaying in wireless powered communication networks (WPCN), where a UAV is used to aid in the transportation of information from numerous sources to a destination with severely obstructed communication channels. The sources are low-power and do not have any energy source. The UAV serves as a hybrid access point (AP) that is used both as WPT to power the user devices and as a means of information transmission and reception. In~\cite{Zaidi2019}, the authors consider the use of UAV as a relay to serve users at the cell edge by extending coverage of an existing network using non-orthogonal multiple access~(NOMA) technology, thereby enhancing their QoS.

\subsection{Capacity/ Throughput Enhancement}
The exponential growth in the telecommunication and information technology landscape over the last decade has witnessed a tremendous increase in the amount of user demands and requests for more resources in terms of data traffic. To cope with the situation, macro BSs and small BSs are used to provide coverage to users. However, due to the continuous increase and fluctuation in data traffic at various locations in the network, the dense deployment of only terrestrial small BS networks is no longer sufficient to address this capacity demands. As a result, UAVs have been identified as a potential solution to provide more reliable and effective coverage to users due to their flexibility, adaptability, and quick configuration~\cite{you2021towards, Saad2020}.

The authors in~\cite{sharma2016uav} considered the optimal deployment of UAVs in heterogeneous networks in order to enhance the capacity of the network. To determine the optimal geographic location to deploy the UAVs, a utility function was developed to model the traffic intensity in various parts of the network, after which a heuristic algorithm was proposed to assign the UAVs to their optimal positions. Compared to existing ground-based wireless networks, the suggested model was proven to offer superior capacity, consistency, and extended connectivity.
The work in~\cite{Qingheng2019} considered the deployment of multiple-UAVs to offload traffic from a single terrestrial BS that is heavily loaded in order to maximize the throughput of ground users at the cell edge.

\subsection{Backhauling}
In cellular networks, the backhaul serves as a connection between the BSs and the core network and this connection is usually established using fibre cable, microwave links, etc,. However, in emergency scenarios, where existing backhaul infrastructures have been damaged, or when there is a need to enhance the capacity of existing backhaul links, ad-hoc backhaul connections can be established using UAVs~\cite{Challita2017, Gapeyenko2018}. 
With respect to UAV-enabled backhaul connections, the authors in~\cite{Challita2017} proposed a UAV-enabled wireless backhaul mechanism for ultra-dense networks. Game theory was employed to model the formation of the multi-hop backhaul network comprising multiple UAVs. Then, a heuristic algorithm was proposed to determine the optimal network formation strategy that maximizes data rate and minimizes network delays.
UAVs can also be used for backhauling in high mobility network scenarios such as in high-speed trains where, due to the very high speed of the train, the channel condition is subject to continuous turbulence and instability. In this regard, a backhauling mechanism using the combination of UAVs and free-space optical~(FSO) communications was developed in~\cite{khallaf2021comprehensive}, to enhance the coverage probability of the high-speed train network.

\subsection{Energy Efficiency}
UAVs can also help in enhancing the EE of existing cellular networks. This can be achieved by deploying UAVs to assist heterogeneous networks to enhance their capacity faster than their power consumption, which in turn helps to improve their EE. The authors in~\cite{alsharoa2017energy} demonstrated this by deploying millimeter wave UAVs alongside macro and small BSs in a three-tier heterogeneous network to enhance cell edge users throughput, which also led to the overall improvement in the EE of the network.
UAVs can also be deployed in existing terrestrial cellular networks to serve delay and rate sensitive users while offloading the traffic of lightly loaded small BSs in order to put them into sleep mode. This approach is known as UAV-assisted BS sleeping strategy and results in a significant reduction in the energy consumption of the network as was introduced in~\cite{chang2021energy}.

\section{Types of UAV deployments}\label{sec:deployment}
\begin{figure*}[ht]
	\centering
	\includegraphics[width=0.95\textwidth]{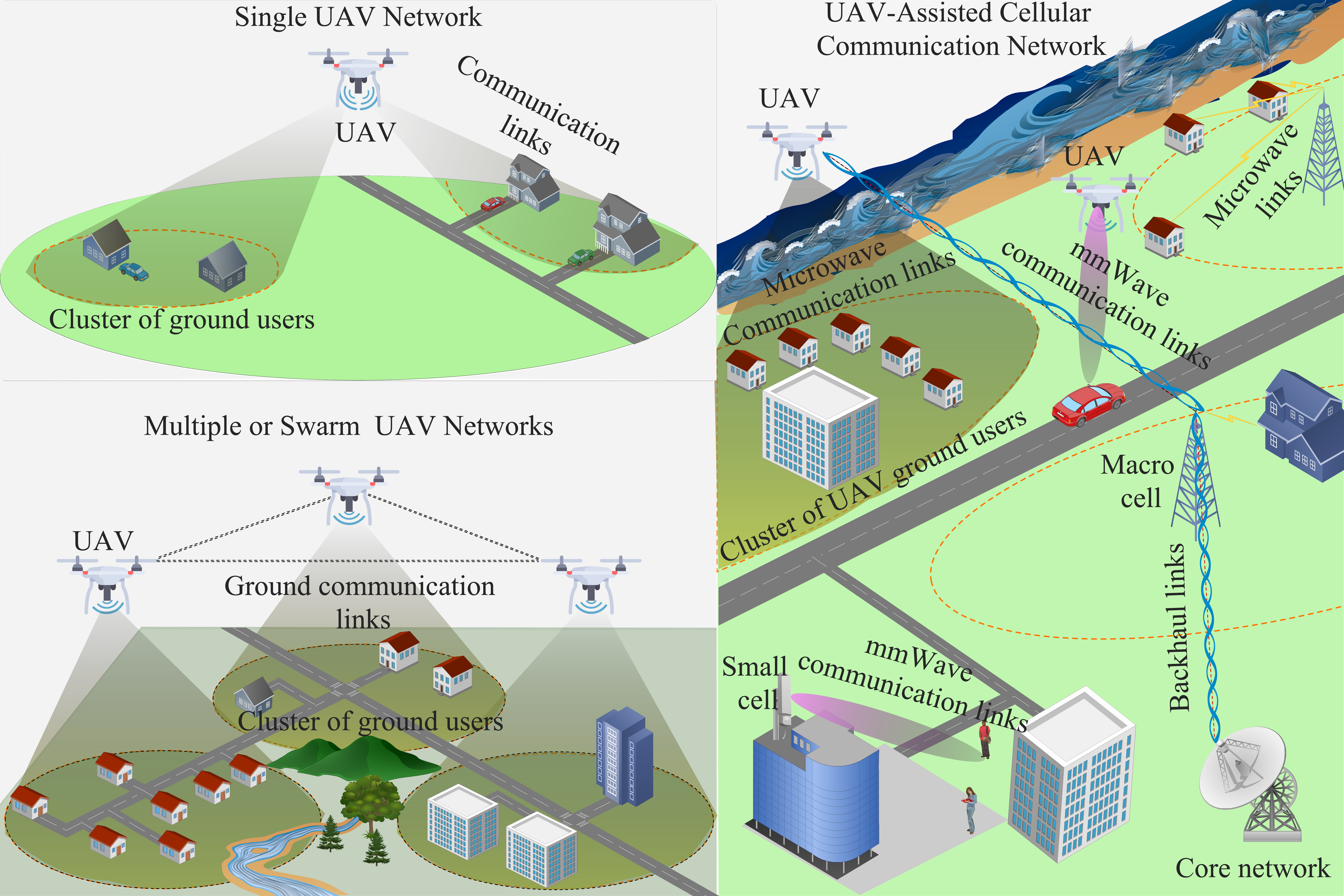}
	\caption{Illustration of the two major types of UAV deployments: UAV stand-alone deployment comprising single and multiple UAV networks, and UAV-assisted cellular networks comprising both UAV-BSs and terrestrial BSs.}
	\label{fig:drone_dep}
\end{figure*}

        As highlighted in Section~\ref{Sec:Four}, there is an ever-increasing application of UAVs in different aspects of wireless communications. In each of these applications, wireless networks can comprise  standalone UAV-BSs, such as in emergency scenarios where, due to natural disaster, the fixed cellular network infrastructures are destroyed, and a pop-up network must be implemented to restore network service in the affected areas~\cite{Deepak2019, paulo-uav}. Wireless networks can also comprise both UAV-BSs and terrestrial BSs. This is the deployment type where the UAV-BSs are deployed to assist existing cellular networks in ensuring service continuity during failure or breakdown of a BS site or to provide coverage and capacity enhancement~\cite{Chakareski2019}. The two major types of UAV-BS deployments are illustrated in Fig.~\ref{fig:drone_dep}, while in the remaining part of this section, we briefly highlight the main features of each type of UAV deployment. 
\subsection{Standalone UAV Deployments}
Standalone UAV deployment involves the deployment of single or multiple UAV-BS to provide network service in an area without fixed cellular network coverage.
Two major deployment scenarios exist in this approach. The first scenario involves the deployment of a single UAV-BS to provide wireless network service in an area, including harvesting data from IoT networks deployed in a field, acting as a relay to provide wireless service to users that have been separated by large obstacles such as mountains and hills, etc.~\cite{ouyang2014optimization}. 
In the second scenario, multiple UAVs are deployed in the form of a swarm network: i) to provide a complete wireless network service to a particular area, which could be a dedicated network for an organization; ii) to provide wireless service in rural areas without prior cellular network infrastructure; iii) to restore network coverage for a large area that has been affected by a natural disaster such as earthquakes or volcanoes~\cite{chen2020review}.
The major challenge with single UAV deployment is that when a fault occurs in the UAV-BS, it could result in complete network failure. On the other hand, in a multi-UAV-BS system, when a single UAV fails, we can reconfigure the system and still have a sub-optimal solution. However, this does not mean that the use of multiple UAV-BSs does not have disadvantages, as the problem of proper coordination among the multiple UAVs deployed exists~\cite{hentati2020comprehensive}.
In summary, choosing between a single UAV or multiple UAVs deployment depends on the nature of the communication system that is being developed and the problem that the network would address.


\subsection{UAV Deployment with Fixed BSs~(UAV-Assisted Cellular Networks)}\label{sec:optimization}
This involves the deployment of either single or multiple UAV-BSs on top of existing fixed cellular network infrastructure~(terrestrial BSs) to provide enhanced capacity and coverage in different scenarios~\cite{alsharoa2017energy}. For example, during major events, such as football matches and concerts involving a large gathering of people, the networks in such areas would be very congested, which could lead to poor QoS. In such scenarios, UAV-BSs can be deployed to provide additional capacity in order to reduce network congestion and to enhance the data rate of the users in hotspot regions and at the cell edges~\cite{Chakareski2019}. 
In addition, UAV-BSs can help to provide communications with highly mobile user equipment~(UE), thereby preventing frequent handovers, which can negatively affect their QoS~\cite{zeng2016wireless}. One of the main advantages of deploying UAVs in existing cellular networks is that they can change their locations depending on the network conditions, which helps in increasing the QoS, restoring the network at failed BS sites, achieving load balancing, offloading traffic from macro BSs, and extending the coverage. Hence, UAV-BS deployment in cellular networks is a key design consideration in future heterogeneous wireless networks that can enable applications such as smart cities, mobile computing,  autonomous vehicle networks, etc.~\cite{bas}.

\section{Energy Optimization in UAV-based cellular networks}\label{sec:EnergyOpt}
The energy consumption of UAVs is one of the major challenges that limit their applications in many areas including in wireless communications networks~\cite{gree-uav,  zeng2016wireless}. As 5G and beyond networks target drastic improvement in network EE, the use of UAV-BSs, though very promising, could be a hindrance to the actualization of this objective if the issue of increased energy consumption is not carefully considered~\cite{Fotouhi2019}.
Since UAVs are designed to fly from one location to another, their application in cellular networks brings a lot of flexibility and adaptability to the network as they can be deployed on demand to various parts of the network to handle different challenges ranging from network restoration, coverage and capacity enhancement, traffic offloading, load balancing, backhauling, etc., as mentioned in Section II.

However, the large amount of energy consumption involved in flying or hovering the UAV to or over the service area makes their use very challenging. 
Although various UAV power supply and charging mechanisms have been discussed in Section III, it is still very important to optimize the energy consumption of the UAV-BSs because of the limitations of these methods including low energy storage capacity, need for frequent recharging, low energy conversion efficiency, unpredictability of renewable energy sources, etc., which can negatively impact the performance of UAVs in wireless networks~\cite{Galkin2019}. In addition, most energy optimization techniques are easier and less expensive to implement compared to the various power supply and recharging techniques discussed in Section III.
Therefore, in this section, we highlight the major areas of energy optimization, while in subsequent sections, we review the various algorithms that have been proposed for tackling the energy optimization problems identified in this section.

There are four major areas of energy optimization in UAV-based cellular networks:
\subsection{Optimization of the Propulsion Energy}
The propulsion energy is the energy consumption associated with flying or hovering the UAV-BSs over the service area. This is the most significant energy consumption of the UAV-BS~\cite{Fotouhi2019}. Hence, energy optimization strategies proposed in this direction aim at reducing the energy consumption due to the movement of the UAV-BS. In this regard, a few models to quantify the energy consumption of the UAV due to propulsion have been developed in~\cite{Zeng2019, Zeng2017, Abeywickrama2018}. In addition, various approaches to minimize the propulsion energy consumption have been introduced, including regulating UAV altitude and planning the trajectory, etc~\cite{Zeng2017, Sambo2019, Weiwei2021}. 
 
\subsection{Optimization of the Communication Energy} 
Communication energy is the energy consumption associated with signal processing and data transmission during the UAV operation in wireless communication networks. This is usually less significant compared to the energy consumption due to propulsion~\cite{Fotouhi2019}, thereby energy optimization approaches proposed in this area is targeted at reducing the energy consumed while processing and transmitting user information. The techniques proposed in this area include transmission power allocation and control, scheduling the transmission of UAV-BSs---particularly for the cases where they need to fly over a predefined trajectory---, and optimally positioning UAVs in a service area~\cite{Alzenad2017, wang2020uav, You2020, Plachy2020}. 

\subsection{Joint Optimization of the Communication and Propulsion Energy} 
Unlike the first and second cases that consider optimizing either the communication or propulsion energy consumption, the works carried out in this direction considers the simultaneous optimization of both the communication and propulsion energy consumption of UAV-BSs. These approaches result in the most energy conservation as both components of the UAV-BS energy consumption are considered together. The strategies considered in this area involve a combination of the approaches proposed in propulsion energy optimization as well as that of communication energy optimization~\cite{Faqir2017, Zeng2019}. 

\subsection{Optimization of the Energy Consumption in UAV-Assisted Cellular Networks}
The previous cases~(i.e., propulsion and communication energy) considered mainly deal with the energy consumption of only the UAV-BSs when they are deployed alone or as a swarm network comprising multiple UAV-BSs.
The UAV-assisted cellular network is the case where a single/multiple UAV-BSs is/are deployed in existing terrestrial BSs in order to enhance certain network performances such as throughput, coverage, etc.~\cite{li2018uav,shehzad2021performance}.
However, their deployment could result in an increase in the overall energy consumption of the network if not properly managed. Hence, the strategies developed in this direction are meant to reduce the energy consumption of both the UAV-BSs and the fixed BSs, or a scenario where the UAV-BSs can help in reducing the energy consumption in fixed BSs through UAV-assisted BS sleeping strategy~\cite{chang2021energy, Chakareski2019, alsharoa2017energy}. 

\section{Overview of Algorithms for Energy Optimization UAV-based Cellular Networks}\label{sec:algorithms}
To ensure energy efficient deployment, operation and management of UAVs, whether as a standalone UAVs network or UAV-assisted cellular network, there is a need for energy optimization strategies to be devised that would ensure that the UAV minimize their energy consumption while serving user demands due to the limited energy capacity of battery-operated UAVs.
However, the development of energy-efficient solutions requires the application of conventional or traditional algorithms and ML algorithms.
Hence, in this section, we review some common algorithms that are used in the development of energy-efficient strategies to minimize the energy consumption of mobile cellular networks.
Then, in section VIII, we consider specific strategies that have been devised to reduce the energy consumption in UAV-based cellular networks alongside the specific algorithms that were applied to achieve the proposed energy-efficient solutions.

\subsection{Conventional Energy Optimization Algorithms}
\begin{figure*}
    \centering
    \includegraphics[width=\textwidth]{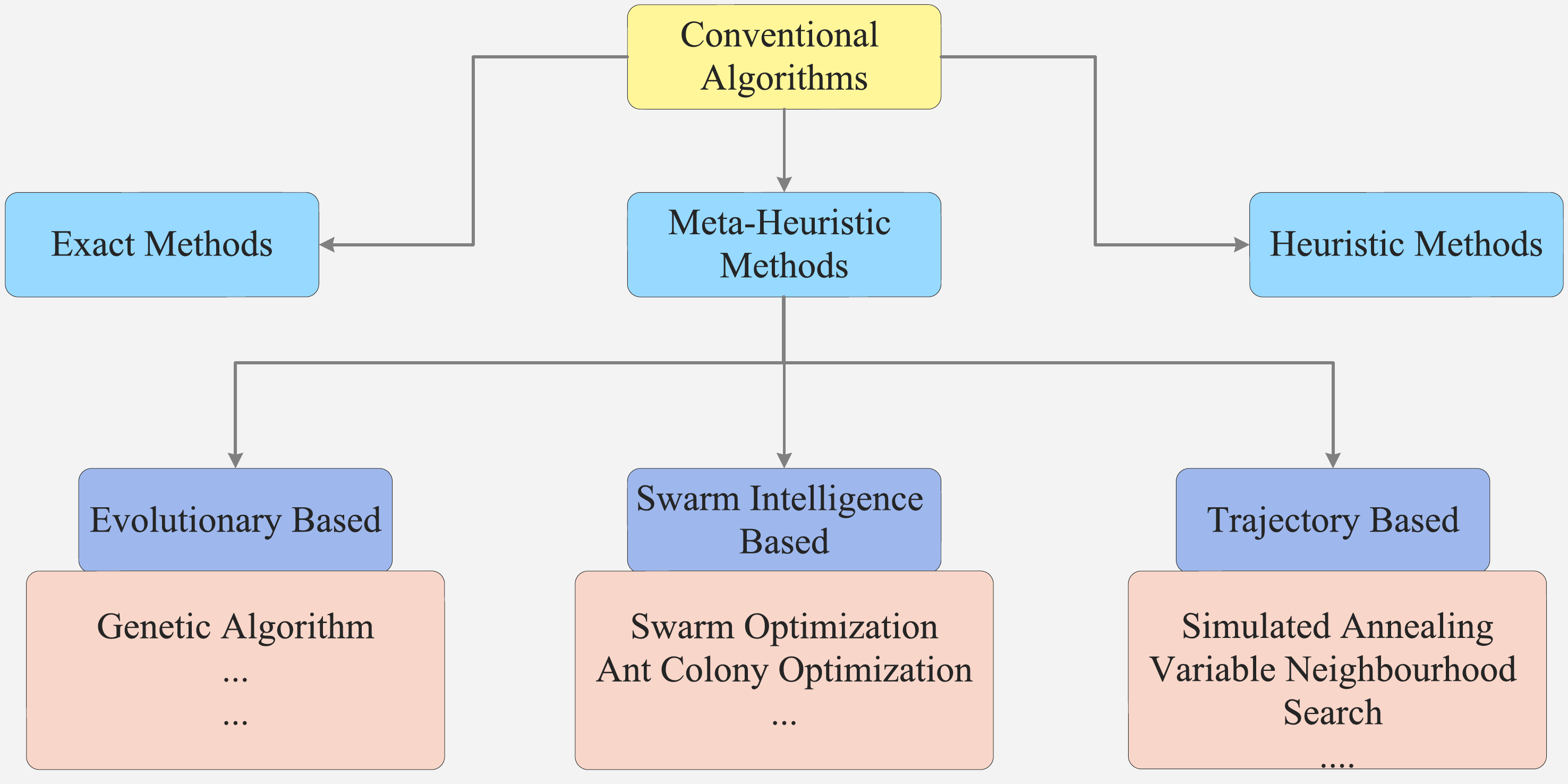}
    \caption{Classification of the conventional algorithms discussed within the scope of this work.}
    \label{fig:conv_algo_map}
\end{figure*}

There are several conventional optimization methods that have been applied for energy optimization in UAV-based cellular networks in the literature.
Conventional methods are divided into three categories: exact methods, heuristic and meta-heuristic methods.
Exact methods are not sufficient, especially in operational decision processes, due to the unacceptable solution times and their inability to reach solutions in large-size problems in a reasonable time~\cite{rothlauf2011optimization}.
Therefore, in this survey, we focus on the remaining two categories: heuristics and meta-heuristics.
Heuristic algorithms are a set of procedures that a developed to  specifically address an optimization problem \cite{cihat_35, cihat_36}.
Meta-heuristic methods can be classified in terms of various features such as neighborhood structure and searching strategies.
In this study, meta-heuristic methods are discussed in terms of three basic classes, namely: evolutionary-, swarm intelligence-, and trajectory-based.
However, since there are many developed meta-heuristic methods in the literature in each class, the most-known methods are examined in a general framework. The general classification of conventional algorithms can be seen in Fig.~\ref{fig:conv_algo_map}. 
In the following paragraphs, we briefly discuss the common conventional algorithms.

\subsubsection{Heuristic Algorithms}
Exact algorithms are developed in such a way that the optimal solution can be achieved in a limited time. However, for some complicated  optimization problems (e.g., NP-hard or global optimization), this limited amount of time may expand exponentially in relation to the problem sizes. Heuristics lack this guarantee and, as a result, often provide solutions that are less than optimum or approximate solutions. Heuristic algorithms frequently find acceptable solutions in a reasonable amount of time. In addition, they are often problem-dependent, that is, they are designed for a specific problem such as energy optimization in cellular networks, UAV routing problems, etc.
There are two heuristic approaches to solving hard optimization problems. One of such approaches is the constructive heuristics which develops solutions via iterations. It is called a constructive heuristic because it begins with an empty solution and continues to expand on it until a complete solution is discovered~\cite{cihat_1}. The other heuristic method is called improvement or local search heuristics. Improvement heuristics start with a complete solution and then strives to improve on the existing solution further by local searches. Examples of heuristic algorithms include block coordinate descent (BCD), Dinkelbacks method, successive convex approximation (SCA), circle parking theory, etc. 

\subsubsection{Meta-Heuristic Algorithms}
Many heuristic algorithms are quite customized and problem specific.
A meta-heuristic, on the other hand, is a high-level problem-independent algorithmic framework that offers a set of principles or techniques for the development of heuristic optimization algorithms. However, a specific definition is quite tricky, and many scholars and practitioners use the terms heuristics and meta-heuristics interchangeably. As a result, the word meta-heuristic may also refer to a problem-specific implementation of a heuristic optimization algorithm based on the rules given in such a framework. Unlike heuristic methods, a meta-heuristic knows nothing about the problem it will be applied. It can treat functions as black boxes. For more information on meta-heuristic algorithms, the studies in~\cite{cihat_38} can be examined. In the following paragraphs, we examine meta-heuristic algorithms in three categories.

\paragraph{Evolutionary-based Algorithms} 
These are algorithms that are derived from the concept of biological evolution. In this regard, the genetic algorithm~(GA) is presented as a representative of the evolutionary-based algorithms, as it is one of the most commonly used for energy-efficient UAV-based communications.
\begin{itemize}
    \item \textbf{Genetic Algorithm:} GA is a meta-heuristic optimization method based on the principles of the biological evolution process that finds the best solution to problems that are difficult to solve with exact methods. The first studies on this algorithm were conducted by John Holland. Holland developed new methods for computer systems based on the principles of natural selection and adaptation existing in nature. Holland argued that processes such as crossover, mutation and selection that take place in the evolutionary process are very important for solving optimization problems and that better individuals can be obtained in each generation. Holland modeled all these processes for solving problems by considering the perfect adaptation of living things in nature to the ecosystem~\cite{cihat_2}.
    
GA is one of the population-based algorithms used for solving complex problems because it provides a convenient and fast solution. The population consists of individuals that make up the solution set. By eliminating the bad solutions in the solution sets created in each generation, the next generations consist of good solutions that will lead to better results. Since there is more than one solution set in a generation, finding many best solutions in one step is one of the features that distinguish GA from other algorithms. Also, by focusing on the part of the solution set, it can do an effective search and provide the best solution in a short time~\cite{cihat_3}.

In the GA application process, the first step is to define how to encode the solutions represented by chromosomes according to different problems. After the necessary parameters are received from the user, the initial population is created so that the GA steps can begin. Each chromosome is an individual and consists of genes. The initial population consists of randomly selected chromosomes. Then, the fitness function that defines the problem solution is determined. Afterwards, chromosomes that will form the next generation are selected from the population, and genetic operators based on genetic processes in nature are applied respectively to obtain better chromosomes. The crossover process is applied to generate new offspring from the individuals obtained from the selection process. The mutation tool is used after the crossover step to provide diversification in the population. At the end of all these processes, new generations are created and compared with the fitness values of other generations.
Individuals with good fitness are preserved and passed on to other generations (elitism). This process continues until a specified termination criterion is met~\cite{cihat_37}.

There are several features that make the GA different from other conventional heuristic methods. The most important of these are that GA offers more than one solution and needs less information for the obtained solutions. Also, GA uses probabilistic transitions rules and can be parallelized very easily for application in both continuous and discrete problems. However, the drawback of using GA is that it is difficult to model the problem using the algorithm, and its implementation involves a high computational cost compared to that of other conventional heuristic approaches~\cite{cihat_3}.
\end{itemize}

\paragraph{Swarm Intelligence-Based Algorithms} These  algorithms are inspired by nature and designed based on the relationship between living organisms, including ants, birds, bees, flocks of fish, bacterial communities, etc~\cite{angadi2021computational}. In this regard, Particle Swarm Optimization~(PSO) and Ant Colony Optimization~(ACO) and considered in the following paragraphs as  representative of swarm intelligence-based algorithms.
\begin{itemize}
    \item \textbf{Particle Swarm Optimization (PSO) Algorithms:}
    PSO algorithm was introduced by Eberhart and Kennedy in 1995~\cite{cihat_4}. It was developed as a population-based optimization method inspired by the two-dimensional behavioral movement of bird and fish flocks in nature. It has a more straightforward computation method than other traditional optimization methods and does not involve time-consuming complex operations. Therefore, it works faster has shorter computation times and is more preferred~\cite{cihat_5}.

The solution approach of the PSO algorithm is as follows: There is a flock of birds in a region where there is only one food. Birds are randomly placed in this food area and no bird knows where the food is. But at the end of each iteration, they know how close they are to the food. In this case, it is a good decision to follow the bird closest to the food. PSO works according to this scenario and is used to solve optimization problems. Birds trying to find food in solution space are called ``particles" in PSO. Each particle has a fitness value and velocity that enables it to fly. These are calculated using the fitness function. Particles fly out of the problem space, following the optimum particle at each iteration~\cite{cihat_6}.

If there is no specific initial solution generation mechanism for a problem, the PSO is started with a group of random solutions (particle swarm) and tries to reach the global best solution with updates. The first obtained feasible solution value is kept as the best solution and the coordinates of the associated solution are determined. In each local search, this value is kept in memory for later use and is called the “local best solution”. The other best value is the coordinates that provide the best solution ever obtained by all particles in the population. This value is kept as ``global best solution". In each iteration, the local best solution is compared with the global best solution based on the objective function to develop the global best solution.

\item \textbf{Ant Colony Optimization Algorithms (ACO):} ACO is a meta-heuristic technique used for solving optimization problems and works based on swarm intelligence as PSO. It was developed by Dorigo et al. in 1991 and tested on different sizes of Traveling Salesman Problems (TSP). Dorigo named this algorithm the 'Ant System'~\cite{cihat_7}. 

The basis of this technique is the pheromone hormone that ants use in communication. Ants start the foraging process randomly, and when food is found, they secrete the pheromone hormone to show the other ants in the colony the pathway to the discovered food. This hormone is updated by other ants and helps the colony find the shortest path to food. Intense pheromone amount indicates the quality of the path and increases the probability of preference for the use of that path. If the ants encounter any obstacle on the way between the food and the nest, the ant in front of the obstacle cannot continue and they must make a decision for the new direction of the trip. Each of the new direction options is equally likely to be selected. If the ant chooses the shortest path, this path becomes the preferred route according to the pheromone hormone density. However, if the chosen path is not the shortest, the colony route is reconstructed very quickly and the amount of pheromone on the newly chosen path is increased to create a preference for the ants that come later. Considering that each ant releases the same amount of hormone at the same rate on average, the expected situation is that it takes a long time for the colony to recognize the obstacle and choose the shortest path. However, the path selection made by the ants coming from behind, depending on the amount of pheromone, shortens the total time to trip for food~\cite{cihat_8}. 
\end{itemize}

\paragraph{Trajectory-Based Algorithms}
These algorithms employ a single agent that traces a trajectory while moving through the search space to determine the global optimal solution. In the process, a better solution is accepted while a  solution that is not so good may be accepted with a specific probability. Hence, in the following paragraphs, we briefly discuss two common trajectory-based methods: simulated annealing and variable neighbourhood search algorithm.
\begin{itemize}
    \item \textbf{Simulated Annealing Algorithm (SA):} SA is a meta-heuristic algorithm developed by Kirkpatrick et al. in 1983 to solve optimization problems. The SA method is based on the analogy between the annealing process in physical systems that minimizes the energy state of the solids and the solution process in combinatorial optimization problems~\cite{cihat_9}. 

The SA algorithm starts with an initial solution and a relatively high-temperature value to avoid being trapped by the local minimum. At each iteration, the algorithm produces the next solution within the local neighbourhood and the temperature decreases according to specific rules. A new solution that represents the energy level of the system and improves the objective function is always accepted. On the other hand, a workaround proposal that allows for an increase in the temperature of the system or allows a certain degree of divergence/deterioration from the objective function in the system is also accepted. The algorithm is conducted with a new solution if the new solution is accepted and with an existing solution if the new solution is rejected. These processes continue until the termination criterion (number of iterations, the smallest temperature value, etc.) are met.
\item \textbf{Variable Neighborhood Search Algorithm (VNS):} The VNS meta-heuristic was developed by Pierre Hansen and Nenad Mladenovic in 1997~\cite{cihat_10}. The VNS method, which has been continuously developed since its inception and has applications in numerous fields, is a single solution-based, static/dynamic objective function, based on various neighborhood structures (meta-heuristics other than VNS use a single neighborhood structure). Based on the systematic modification of neighborhood structures used in the search, VNS is a simple and effective meta-heuristic aimed at solving combinatorial optimization problems. Since the local minimum in any neighborhood may not be valid for other neighborhoods, the use of the multiple neighborhood structures is advantageous because it enables the best solutions in different regions of the search space to be obtained. In addition, these neighborhood structures are systematically changed during the search process. Thus, by providing diversification in the search space, the disadvantage of being stuck in the local optimum can be overcome. VNS offers significant advantages over other algorithms due to its simple structure, integration with different solution techniques and it requires few parameters.

\end{itemize}

\subsection{Machine Learning Algorithms}
\begin{figure*}
    \centering
    \includegraphics[width=\textwidth]{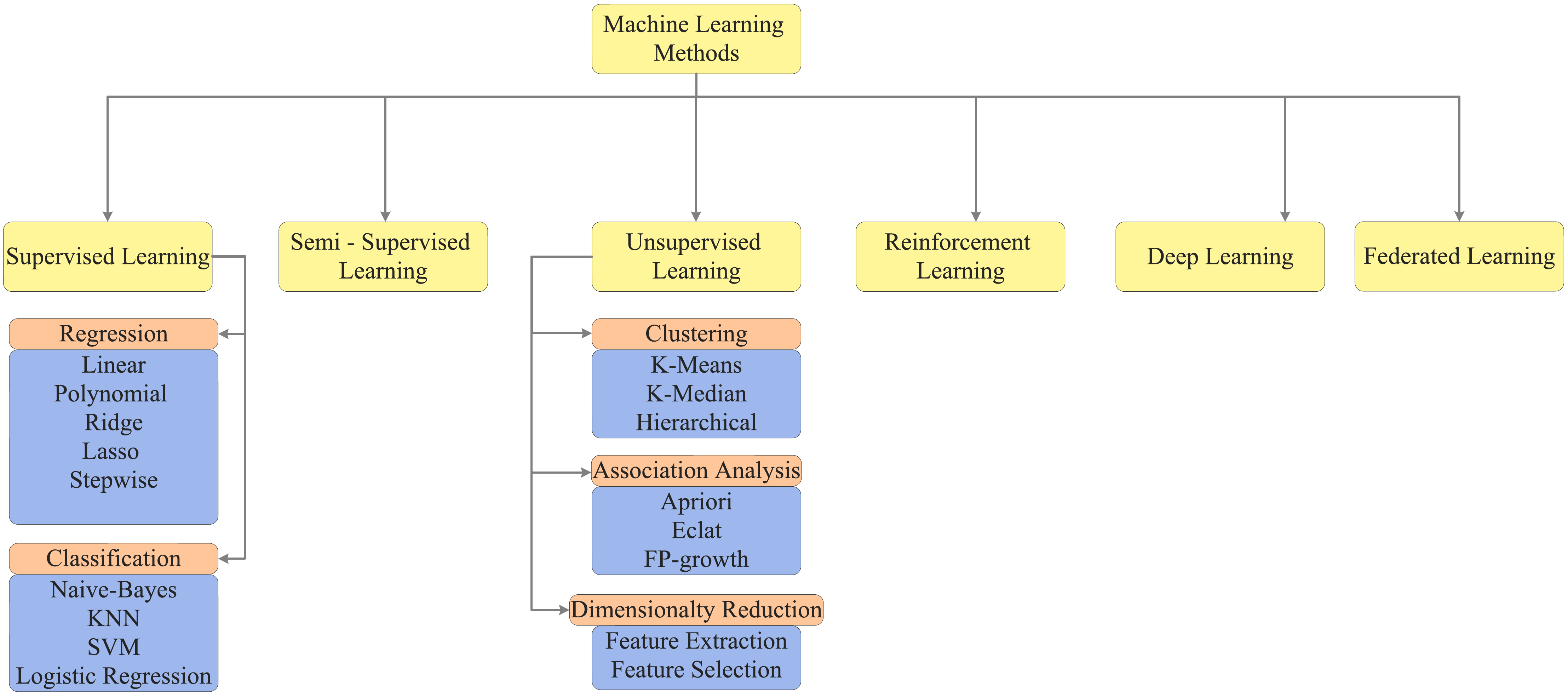}
    \caption{Classification of machine learning algorithms within the scope of this work.}
    \label{fig:ml_algo_map}
\end{figure*}

ML are a class of algorithms that can learn from data without being explicitly programmed~\cite{ChaoyunANN}.
Generally, they can be classified according to the amount and type of supervision they get during their training period. There are six major categories of ML algorithms: supervised learning, unsupervised learning, semi-supervised learning, deep learning~(DL), reinforcement learning~(RL), federated learning~(FL). These categories can be seen in Fig.~\ref{fig:ml_algo_map}. 

\subsubsection{Supervised Learning (SL)}
In SL, the training set comprises the input data set and the desired output, referred to as labels. SL learns a function (a match between the input data and the result data) by extracting information from the input data and the labels fed into the machine~\cite{cihat_11}. SL problems are divided into two main categories, which are regression and classification~\cite{cihat_12}. In regression problems, a continuous output is predicted; that is, the input variables are mapped to a continuous output. Examples of regression SL methods are Linear regression, ridge regression, step-wise regression, etc~\cite{cihat_13}. In classification problems, a categorical output is predicted. That is, the input variables are mapped to different output classes, which could be a binary class~(comprising two classes) or multi-class~(comprising many classes). Logistic regression, naïve Bayes classifier, k-nearest neighbor and support vector machine are examples of  classification SL method~\cite{cihat_12}. 

\subsubsection{Unsupervised Learning~(UL)}
UL algorithms work with a set of inputs. The input data set for training does not have a labeled output. For this reason, in UL, clustering, association, and pattern discovery are performed over existing data. It has a working mechanism that is different from SL. The purpose of UL is to enable the identification of patterns within the training datasets and categorize input objects according to the patterns defined by the system~\cite{cihat_14}. These algorithms are expected to develop specific outputs from unstructured inputs by looking for unexplored relationships between each instance or input object. UL algorithms can be classified into three main groups: clustering, association analysis and dimensionality reduction. K-means, K-median, hierarchical clustering, and expectation-maximization are the most common examples of clustering category~\cite{cihat_15,cihat_16}. APRIORI, Eclat and FP-Growth are examples of association analysis models~\cite{cihat_17}. Principal component analysis~(PCA) and linear discriminant analysis~(LDA) are examples of dimensionality reduction category~\cite{cihat_18}.



\subsubsection{Semi-supervised Learning (SSL)}
SSL is an approach that combines a small amount of labeled data with a large amount of unlabeled data in the training stage. It is a hybrid form of supervised and unsupervised learning with  weak supervision features. A small amount of labeled data can play an important role in improving  learning accuracy. While labeling all training data set is not a feasible solution, labeling some of the data set incurs an acceptable cost. Thus, SSL algorithms can be developed by hybridizing clustering and classification  algorithms. It enables us to obtain the most suitable samples in the data set by collecting the data according to their similarity with the clustering step. Then,  the data obtained from each cluster are labeled and used in the training of the supervised ML algorithm for classification~\cite{cihat_39}.

\subsubsection{Deep Learning~(DL)}
DL is a branch of ML that mimics the functioning of the human brain in the data processing. It allows machines to learn without human supervision. It gives the ability to perceive spoken, translate, identify objects, and make decisions. Despite being a branch of ML, DL systems do not have limited learning capabilities like traditional ML algorithms. Instead, DL systems can continually improve their performance as they are fed larger and more consistent data~\cite{cihat_20}.

DL was realised with the help of artificial neural networks. Numerous sensors, the artificial counterparts of neurons, have come together to form artificial neural networks. The term ``deep" is used to denote the number of hidden layers that neural networks have. Each layer of the neural network processes the input data and transmits it to the next layer until the data reaches the output layer, where the desired output is obtained. DL algorithms have sophisticated architecture comprising in-built feature extraction and representation capabilities  absent in traditional ML algorithms. This enables them to process very complex data such as images, videos, natural language, play games, etc~\cite{cihat_21}.

Another way to understand how DL works is to look at Convolutional Neural Networks (CNNs). Convolutional neural networks detect features directly, eliminating the need for manual labeling of data. None of the features is previously taught to the machine; instead, it trains itself on the given set of images. This automatic feature detection capability makes DL models useful for object classification and other computer vision applications. The reason deep neural networks are so sensitive in identifying features and classifying images is because of the hundreds of layers they contain. Each layer learns to describe certain features, and as the number of layers increases, the complexity of the learned image features changes in direct proportion~\cite{cihat_21,cihat_22,cihat_23}.

\subsubsection{Reinforcement Learning}
Although it is not completely different from supervised  and unsupervised learning methods, RL imitates human's learning process. It shows how a system can perceive its environment and learn to make the right decisions in order to reach its goal. It differs from both supervised and unsupervised learning in that the agent is not given any prior knowledge of the environment, such as input data and output data, but gathers information about the environment by interacting with the environment and learning to take the right action in any given situation (for example, by repeated trial and error over a period). This method is frequently used in fields such as robotics, game programming, disease diagnosis and factory automation~\cite{cihat_24,cihat_25}. 

RL approach consists of some basic elements. One of them is the state, which is a set, S,  containing the state and environment of the agent. Another element is the set of actions, A, which constitutes the whole of actions that the agent can perform. Apart from the S and A sets, there are the principles that determine the agent's actions in the action process and the rules that describe the agent's observations. In this context, an agent interacts with the environment by performing an action. Evaluates the new state and the rewards obtained by observing the environment. Actions and observations follow each other in a cycle. In this cycle, each action gives feedback to the agent. The learning process is normally guided by a reward function which is a measure of how much the action that the agent takes in particular state contributes to archiving the desired goal. There are three main categories of RL, including policy-based. e.g., reinforce algorithm, value based, e.g., Q-learning and SARSA and actor-critic RL~\cite{cihat_26}.

\subsubsection{Federated Learning}
FL is an emerging model that is used for enhancing data privacy in DL through distributed learning and centralized aggregation. This means that in FL, the learning algorithm is trained across multiple decentralized edge devices and servers containing local data samples without exchanging information among the various entities~\cite{cihat_31}. However, FL enables advanced devices to collaboratively train a shared model but without sharing their local data. Confidentiality and security of both central and local parameters are important when developing applications in FL. Different privacy algorithms are used for this. The application of FL enables the development of secure ML models, especially in sectors such as health, education, and banking, where data privacy is important. 

FL can be divided into three different phases. These phases usually start with a basic model that is trained on a central server. In the first step, this generic model is sent to the client devices of the application. These local copies are then trained on the local data generated by the client systems, and the performance of the ML model is improved in the desired direction. In the second step, the parameters learned by the ML models are sent to a public server. These processes occur periodically on a specific schedule. In the third step, the central server collects these parameters. After the parameters are collected, the central model is updated and shared with the clients again~\cite{cihat_28,cihat_29}.

Privacy is the most important advantage of FL. In addition, the reduction or elimination of network delays while working on various devices shows the importance of the FL approach in terms of efficiency. In addition, the costs of sharing the data with the server are eliminated. However, the FL approach has the challenges of synchronizing multiple clients as well maintaining data homogeneity. 
Therefore, to implement FL, ML practitioners need to adopt new tools and a new way of thinking regarding model development, training and evaluation without direct access to or classification of raw data while considering communication cost as a limiting factor~\cite{cihat_30}.

\section{Energy optimization techniques in UAV-based cellular networks}\label{Sec:eight}
Energy optimization is very important in UAV-based cellular networks because it prolongs the battery lifetime of the UAV-BSs thereby reducing the OPEX; it minimizes greenhouse gas emissions; and it also reduces the CAPEX as it enables the deployment of fewer UAVs~\cite{Fotouhi2019, mozaffari2019tutorial}. Hence, as the next generation of cellular networks envisions the adoption of both airborne and ground-based BSs in a 3-dimensional deployment scenario, there is a need to more carefully consider the energy consumption of UAV-BSs in order to prevent their energy consumption from escalating. This would not only result in reduced electricity bills for mobile network operators but, more importantly, would lead to a reduction in greenhouse gas emissions to enable both government and corporate organizations to achieve the net-zero emission drive~\cite{gree-uav}. 

In Section VI, we considered the four major categories of energy optimization in UAV-based cellular networks, while in Section VII, we presented an overview of some of the popular algorithms, both conventional and ML, that can be applied for energy optimization in UAV-based cellular networks. This section reviews the specific approaches proposed in the literature for energy optimization in UAV-based cellular networks and the specific algorithms used in implementing such approaches. Hence, for each deployment type, standalone UAV networks and UAV-assisted cellular networks, we first identify various approaches for energy optimization. Then under each approach, we review the specific conventional and ML algorithms used to minimize the network's energy consumption.

\subsection{Energy Optimization of Standalone UAV Deployments}
In this subsection, we consider the approaches that deal with optimization of the energy consumption of only the UAV-BSs, without considering the effect of UAV deployment on the overall energy consumption of the network. 
In this regard, only the communication or propulsion or both energy consumption components of the UAV-BS are considered. In addition, both single and multiple UAV-BS(s) deployments are considered as well as the various algorithms that have been proposed.

\subsubsection{Positioning and placement}
Here, we review the conventional and ML algorithms used for optimal positioning or placement of the UAV-BS in order to minimize their energy consumption. The positioning includes optimizing the location, the altitude, and the radius of coverage of the UAV-BS to maximize EE while ensuring QoS of the ground users. We consider the positioning of a single/multiple UAV-BS/BSs.

\paragraph{Conventional Approaches}
In~\cite{Alzenad2017}, the authors considered the problem of energy efficient 3D-placement of a UAV-BS for coverage maximization. The problem was first modeled as a circle placement problem and a heuristic algorithm was used to determine the optimal 3D location that maximizes the coverage area while minimizing the transmit power. The work in~\cite{Babu2021} investigated the cost and energy optimization of a UAV-based communication network while considering both the communication and propulsion energy consumption. In this regard, a multi-level circle parking~(MCP) algorithm was developed to determine the optimal 3D-hovering positions of the UAVs that maximizes both the uplink and downlink global EE of the network. In addition, the result of the optimal hovering positions obtained was used to determine the number of UAV-BSs and flight parameters required to minimize the total system cost.
The authors in~\cite{Gao2019} proposed a deployment decision mechanism for optimizing the number and locations of UAV-BSs in a UAV-assisted vehicular network to maximize the communication coverage and minimize the energy consumption of the UAV-BSs. The proposed mechanism employs circle packing theory to determine the optimal positions of the UAV-BSs, while an energy optimization model was developed to minimize the power consumption of the UAV-BSs.

The authors in~\cite{Mozaffari2015} developed an analytic solution to determine the optimal altitude for a UAV-BS whereby the transmit power needed to provide coverage to a specific area is minimized.
In~\cite{Khalil2020}, an EE maximization approach was proposed for a UAV-BS relay system to extend the battery life while maintaining network throughput. In the proposed approach, the hovering position, where the UAV-BS expends the least energy, is considered to be the optimal UAV-BS location, is determined via mathematical analysis, after which the power allocation was also optimized. 
The work in~\cite{Nithin2020} considered the optimal positioning of a UAV-BS in order to maximize its EE with the altitude and minimum user data rate being constraints. The EE problem was formulated as a monotonic fractional optimization problem and solved using polyblock outer approximation algorithm.
Two UAV location optimization algorithms were proposed in~\cite{Wang2020} to minimize the transmit power of the UAV-BS. The first algorithm assumes equal power allocation, while the second algorithm is based on successive convex approximation~(SCA) and does not assume equal power allocation. 

The authors in~\cite{Plachy2020} proposed a UAV-BS positioning algorithm based on Coulomb's law to maximize the EE of the UAV-BSs while considering interference between UAV-BSs and user requirements.
An energy-aware 3D deployment algorithm based on Lagrangian and sub-gradient projection for optimal placement of the UAV-BSs was proposed in~\cite{Chou2020}.
In~\cite{Bera2020}, the authors developed a framework for optimizing the energy consumption of individual UAV-BSs in a multiple UAV-BSs network while carrying out location specific tasks. The proposed framework uses order-K Markov predictor to estimate the task locations to enable proactive deployment of UAV-BSs and minimize their energy consumption. In addition, a heuristic algorithm was developed to place the UAV-BSs in their right locations as well as assign their respective tasks to them.
The authors in~\cite{You2020} investigated the optimal 3D placement of a UAV-BS with a tilting antenna to provide sufficient coverage for ground users while utilizing minimum energy consumption. A gradient descent algorithm was then proposed to find the optimal altitude of the UAV-BSs.

The authors in~\cite{Lu2017} considered the importance of the on-board circuit power consumption of the UAVs while addressing the problem of their optimal 3D placement in order to maximize the network lifetime. Then, using an analytical approach, the optimal hovering altitude of the UAV-BSs with respect to their coverage radius was derived to determine the coverage and on-board circuit power parameters that result in minimum power consumption.
The work in~\cite{Nit2021} considered the energy efficient placement of UAV-BSs for data collection from ground users based on NOMA. A heuristic algorithm was proposed to determine the optimal hovering height of the UAV-BS that maximizes the EE of the network. 
The authors in~\cite{Xue2018} proposed a joint 3D location and transmit power optimization scheme for UAV-based relay networks to maximize the sum rate of users.  A heuristic algorithm based on alternating descent and SCA was developed to solve the optimization problem.  

The authors in~\cite{Shakoor2021} proposed a joint optimization scheme for both the 3D placement and pathloss factor with the aim of achieving maximum energy efficient coverage. A heuristic algorithm was developed to find the optimal UAV placement and compensation factor that maximizes the energy efficient coverage.
An optimal UAV placement framework that aims to find the optimal UAV locations that is required to minimize the total energy consumption of the network while providing a target coverage was introduced in~\cite{zorbas2016optimal}. Both centralized and localized heuristic algorithms were developed to determine the optimal UAV locations for both static and mobile users.
The authors in~\cite{Jiang2020} considered the joint optimization of the transmission power and location of UAV-BS in a relay NOMA network to minimize the power consumption of the network. A double loop iterative algorithm was developed to solve the joint optimization problem. 
In~\cite{Bahr2020}, the optimal 3D placement for UAVs serving as relays in IoT communications was considered in order to minimize the transmission power of the UAVs while considering the outage probability of the IoT devices. A 3D placement algorithm based on PSO was developed to minimize the transmitted power in both air-to-ground and ground-to-air links.

The case of energy efficient UAV placements in indoor environments for emergency wireless coverage was considered in~\cite{Cui2018}. Both iterative and exhaustive search algorithms were developed to determine the optimal position of the UAV in order to minimize the transmission power.
Similarly, the authors in~\cite{Pandey2019} investigated the optimal positioning of a UAV-BS for seamless IoT connectivity in an indoor environment comprising multiple users at random locations in order to minimize the transmit power of the UAV-BS. An energy efficient low complexity heuristic algorithm was developed to solve the optimal UAV placement problem.
The authors in~\cite{Bozkaya2020} proposed a UAV-BS deployment and scheduling mechanism to ensure optimal placement and effective management of UAV-BS operations while minimizing energy consumption and ensuring maximum coverage. To achieve these objectives, heuristic algorithms were proposed to ensure the UAVs are placed in the right locations as well as manage their battery recharging cycle. 

The authors in~\cite{Sohail2019} investigate EE maximization in UAV-assisted NOMA based network via a joint optimization of UAV placement and power allocation while considering QoS constraints. The joint optimization problem was modeled as a non-linear fractional problem, then an alternating algorithm based on a nested Dinkelbach structure was proposed to find the optimal solution.
The work in~\cite{Zhang2021} studied the joint optimization of the UAV location and transmit power in a NOMA-based UAV network while considering the decoding order. The joint optimization problem was first divided into two sub-problems after which an iterative algorithm was proposed to solve the optimization problem alternately.
The authors in~\cite{pang2021} proposed an energy efficient transmission mechanism for UAV-enabled mmWave communication system with NOMA by jointly optimizing the UAV position, power allocation, and precoding in order to maximize user coverage and minimize the energy consumption of the UAVs. 
Due to the complexity of the optimization problem, it was first divided into three sub-problems, and three heuristic algorithms were designed to solve each problem iteratively.

\paragraph{Machine Learning Approaches}
The authors in~\cite{Cheng2021} proposed a proactive power control and positioning framework for UAV-BSs to minimize interference and enhance EE in multi-UAV systems. The proposed framework comprises both offline and online phases. In the former, a supervised learning algorithm~(random forest) leverages historical data to build a mobility prediction model, while in the latter, the predicted user positions are exploited to determine the sleep/wake status of the DSCs proactively, while an unsupervised ML algorithm~($k$-means) is employed to update DSCs positions and regulate the power consumption.
An energy efficient multi-UAV deployment framework was proposed in~\cite{Noh2020} in order to maximize user coverage probability. An ellipse clustering algorithm was developed to determine the optimal hovering altitude of the UAV that would result in minimal transmit power while maintaining QoS constraints.

A predictive on-demand ML-based UAV deployment for minimizing both the communication and propulsion energy consumption was introduced in~\cite{Qianqian2018}. In this regard, a ML framework was developed, which uses Gaussian mixture model~(GMM) and weighted expectation maximization~(WEM) algorithm to forecast the network traffic congestion areas. Then, $k$-means algorithm was used to partition the service area of each UAV, after which a gradient-based algorithm was developed to determine the optimal location of the UAVs that results in minimum energy consumption.
The authors in~\cite{Liu2021} considered the problem of reducing the energy consumption required to provide coverage in a multiple UAV network. In pursuit of this objective, a coverage model based on the actor-critic RL algorithm was developed to enhance the cooperation of the UAVs in order to provide the energy efficient coverage. Table~\ref{tab:UAV_pos} summarizes the conventional~(CA) and ML approaches that are applied for optimal UAV-BS placement or positioning in order to minimize its energy consumption.

\begin{table}
\centering
\caption{Summary of energy optimization techniques based on optimal UAV placement and positioning}
\label{tab:UAV_pos}
\begin{tabular}{@{}llclc@{}}
\toprule
\multirow{2}{*}{\textbf{Paper}} & \multirow{2}{*}{\textbf{Year}} & \multicolumn{2}{l}{\textbf{Category}}                            & \multirow{2}{*}{\textbf{\begin{tabular}[c]{@{}c@{}}Specific\\ Algorithm\end{tabular}}} \\ \cmidrule(lr){3-4}
                                &                                & \multicolumn{1}{l}{\textbf{CA}} & \textbf{ML}                    &                                                                                        \\ \midrule
\cite{Alzenad2017}              & 2017                           & \checkmark                      &                                & Heuristic                                                                              \\ \midrule
\cite{Babu2021}                 & 2021                           & \checkmark                      &                                & Heuristic                                                                              \\ \midrule
\cite{Gao2019}                  & 2019                           & \checkmark                      &                                & Heuristic                                                                              \\ \midrule
\cite{Nithin2020}               & 2020                           & \checkmark                      &                                & Heuristic                                                                              \\ \midrule
\cite{Wang2020}                 & 2020                           & \checkmark                      &                                & Heuristic                                                                              \\ \midrule
\cite{Plachy2020}               & 2020                           & \checkmark                      &                                & Heuristic                                                                              \\ \midrule
\cite{Chou2020}                 & 2020                           & \checkmark                      &                                & Heuristic                                                                              \\ \midrule
\cite{Bera2020}                 & 2020                           & \checkmark                      &                                & Heuristic                                                                              \\ \midrule
\cite{You2020}                  & 2020                           & \checkmark                      &                                & Heuristic                                                                              \\ \midrule
\cite{Lu2017}                   & 2017                           & \checkmark                      &                                & Heuristic                                                                              \\ \midrule
\cite{Nit2021}                  & 2021                           & \checkmark                      &                                & Heuristic                                                                              \\ \midrule
\cite{Xue2018}                  & 2018                           & \checkmark                      &                                & Heuristic                                                                              \\ \midrule
\cite{Shakoor2021}              & 2021                           & \checkmark                      &                                & Heuristic                                                                              \\ \midrule
\cite{zorbas2016optimal}        & 2016                           & \checkmark                      &                                & Heuristic                                                                              \\ \midrule
\cite{Jiang2020}                & 2020                           & \checkmark                      &                                & Heuristic                                                                              \\ \midrule
\cite{Bahr2020}                 & 2020                           & \checkmark                      &                                & PSO                                                                              \\ \midrule
\cite{Cui2018}                  & 2018                           & \checkmark                      &                                & Heuristic                                                                              \\ \midrule
\cite{Pandey2019}               & 2019                           & \checkmark                      &                                & Heuristic                                                                              \\ \midrule
\cite{Bozkaya2020}              & 2020                           & \checkmark                      &                                & Heuristic                                                                              \\ \midrule
\cite{Sohail2019}               & 2019                           & \checkmark                      &                                & Heuristic                                                                              \\ \midrule
\cite{Zhang2021}                & 2021                           & \checkmark                      &                                & Heuristic                                                                              \\ \midrule
\cite{Cheng2021}                & 2021                           & \multicolumn{1}{l}{}            & \multicolumn{1}{c}{\checkmark} & Random forest, $k$-means                                                               \\ \midrule
\cite{Noh2020}                  & 2020                           & \multicolumn{1}{l}{}            & \multicolumn{1}{c}{\checkmark} & Clustering algorithm                                                                   \\ \midrule
\cite{Qianqian2018}             & 2018                           & \multicolumn{1}{l}{}            & \multicolumn{1}{c}{\checkmark} & GMM, WEM, $k$-means                                                                    \\ \midrule
\cite{Liu2021}                  & 2021                           & \multicolumn{1}{l}{}            & \multicolumn{1}{c}{\checkmark} & Actor-critic RL                                                                           \\ \bottomrule
\end{tabular}
\end{table}

\subsubsection{Trajectory Design and Path Planning}
Here, the UAV's flight path and speed are carefully planned and designed to ensure that it flies through the optimal trajectory, with the optimal speed and altitude that is required to serve ground users' demands while expending the minimum amount of energy.
In the reviewed studies, energy optimization might be the main objective, such that it can be one of the variables that were optimized or one of the constraints that is considered while optimizing other variables.
\paragraph{Conventional Approaches}
In~\cite{Song2020}, the authors proposed a flight and communication protocol where the UAV-BS's speed, height, and beamwidth are jointly designed to minimize the flight time and energy consumption while serving ground users.
The authors in~\cite{Zeng2017} investigated energy efficient UAV-BS's communication with ground users based on trajectory optimization by considering both the throughput and energy consumption of the UAV-BS. A propulsion power consumption model for fixed-wing UAVs was first introduced, then both unconstrained and constrained trajectory optimization was considered for energy efficient UAV-assisted communications. The former trajectory optimization was proven to be energy inefficient while, for the latter, an energy efficient algorithm based on both linear state space and sequential convex optimization was developed to optimize the communication of the UAV-BS.

The work in~\cite{Zeng2019} considered energy efficient UAV communication of rotary-wing UAV-BSs. A theoretical model of the mobility power consumption was first derived, followed by formulating the UAV energy minimization problem, which jointly optimizes its trajectory, time spent communicating with ground users, and mission completion time. To solve this optimization problem, two solution approaches were considered. In the first approach, a flight-hover communication design was considered where the UAV-BS only visits the optimal hovering location and communicates with the user in that location. As such, the energy minimization problem becomes that of finding the optimal hovering location, duration, visiting order, and speed; and a convex optimization with travelling salesman algorithm was developed to tackle this problem. For the second solution approach, the energy minimization problem was modeled as a path selection problem, then an iterative algorithm based on SCA was developed to update both the trajectory and time allocated to the UAV-BS for communication.
EE of UAV communication based on trajectory design was studied in~\cite{Ahmed2020}. The optimization problem comprises both the throughput and propulsion energy consumption of the UAV-BS. A heuristic algorithm based on SCA and the classical Dinkelbach approach was developed to find the optimal energy efficient strategy for the UAV-BS.

The authors in~\cite{Tianyu2020} considered a joint optimization of the 3D trajectory and transmission power of a multi-UAV based relaying system in order to maximize user throughput. A heuristic algorithm based on the block coordinate ascent~(BCA) technique was proposed to solve the developed problem.
In~\cite{Tran2020}, the authors considered energy minimization problem in UAV based networks by jointly optimizing the trajectory and velocity of the UAV while respecting latency constraints. First, both heuristic and dynamic programming algorithms were proposed to determine the optimal set of trajectories, after which the UAV velocity of each trajectory was also optimized to ensure that the total energy consumption of the network is minimized.
The authors in~\cite{Jing2021} investigated energy-aware trajectory optimization that minimizes the on-board energy consumption in the UAV-BS while satisfying the data requirement of ground users. A double loop iterative algorithm was proposed to determine the optimal trajectory of the UAV that would maximize user coverage.

A combined optimization of both the trajectory and transmit power of a UAV relay network was carried out in~\cite{Zhang2018}. They first derived an analytic expression for the outage probability, after which a heuristic algorithm was developed for power control and trajectory optimization.
The work in~\cite{Sambo2019} investigated the application of UAV-BSs for backhaul connection to terrestrial BSs during post-disaster scenarios. An energy efficient trajectory design based on GA was proposed to select the optimal trajectory that results in the least energy consumption in the UAV.
The trade-off in energy consumption between ground terminals and UAV-BSs was studied using trajectory design in~\cite{Yang2018}. Analytical models were developed to determine the optimal UAV trajectory and ground terminal's transmit power, resulting in the optimal energy consumption trade-off between them.
The authors in~\cite{Hua2019} studied the problem of EE and security in a cooperative multi-UAV system. A joint optimization of the transmit power, user schedule, and UAV trajectory was proposed to maximize the secure EE of the network. A three-layer heuristic algorithm was proposed based on BCD and Dinkelbach method to solve the joint optimization problem.

A UAV-BS path planning framework was proposed in~\cite{Shivgan2020} in order to maximize the EE of the UAV-BS. In the proposed framework, the path planning problem was first modeled using the travelling salesman problem after which a GA was developed to determine the optimal path with minimal turns that would result in minimum energy consumption in the UAV-BS.
The problem of energy efficient UAV-based communication in the presence of several jammers was considered in~\cite{Weiwei2021}. To solve this problem, the UAV trajectory was optimized while considering its mobility constraints. In this regard, a heuristic algorithm based on SCA and Dinkelbach's model was proposed to determine the optimal trajectory.
A flight planning mechanism for solving point of interest visiting problems involving data collection, edge computing, and surveillance operations was proposed~\cite{Jianping2021} to minimize the flight energy consumption of the UAV. The flight turning and switching cost were modeled as a graph and then transferred to a travelling salesman problem after which a heuristic algorithm was proposed to determine the optimal flight path. 

The authors in~\cite{Jiang2019} investigated the rate fairness and EE of a UAV based relay system. Two optimization objectives were considered: the first is the rate maximization while maintaining a specific power budget, and the second is optimizing the total power consumption while maintaining minimum rate requirements. Two heuristic algorithms were proposed to solve the optimization problem. 
In~\cite{Haichao2018}, a joint trajectory and transmit power optimization scheme for UAV-aided communications was investigated to maximize user average throughput. A heuristic algorithm was proposed to solve the joint optimization problem while ensuring that users' throughput is maximized.
Similarly, the work in~\cite{Miao2018} considered a joint trajectory and transmit power optimization in UAV-based secure communications to maximize the throughput of the system. An iterative algorithm based BCD, S-procedure and SCA was proposed to solve the optimization problem.
The work in~\cite{Khamidehi2019} investigated the problem of trajectory optimization to minimize the power consumption of the UAV while ensuring that user service demands are met. The optimization problem was solved using an iterative algorithm that is based on SCA. 

The work in~\cite{Hong2020} investigated flight radius adjustment and routing in UAV swarm networks to minimize the total power consumption of a multiple UAV systems. Two flight radius manipulation approaches were first developed to maximize the propulsion power consumption and battery lifetime of the UAVs.  Then, a temporal path algorithm was proposed to minimize the power consumption of any pair of UAVs. In addition, an iterative algorithm was also developed to further adjust the flight radius to enhance the EE of the system.
The authors in~\cite{Xiang2020} studied the dynamic cooperation of UAVs in a multi-UAV system to enable power efficient communications. To achieve energy efficient communications, the UAV trajectory and the cooperative beamforming were jointly optimized and then a low-complexity heuristic algorithm that is based on difference of convex~(DC) was proposed to solve the optimization problem.
The issue of energy consumption fairness in multi-UAV networks, trajectory optimization was considered in~\cite{Lee2020}. The problem of designing trajectories with fair energy consumption allocation was modeled using mixed integer linear programming. To achieve this objective, two heuristic algorithms were proposed to find the optimal energy allocation strategy among the trajectories that would result in minimum energy consumption in the network.  

In~\cite{Gao2021}, the authors considered energy minimization in a UAV-based network with multiple eavesdroppers. The energy minimization problem was first modeled as a mixed integer optimization problem involving the UAV trajectory and user scheduling while maintaining QoS constraints. Then, an iterative algorithm based on BCD and SCA was proposed to find the optimal solution that results minimum energy consumption in the network.
An EE framework via trajectory optimization in a fixed-wing UAV---used for data collection and forwarding between ground nodes---was proposed in~\cite{Dong2019} to minimize the energy consumption while respecting throughput constraints. A heuristic algorithm based on SCA was formulated to determine the optimal strategy that would result in minimum propulsion energy consumption.
The work in~\cite{Bian2020} studied EE and throughput maximization in UAV-enabled vehicular networks via joint optimization of the UAV trajectory design and power allocation. The joint optimization problem was first divided into two, and then a heuristic algorithm based on SCA was developed to simultaneously determine the optimal trajectory and power allocation strategy in each iteration. In addition, an EE optimization algorithm was developed based on fractional programming and sequential optimization.

In~\cite{Eom2020}, the authors considered the joint optimization of UAV trajectory, velocity, acceleration and transmit power to maximize the EE while maintaining the minimum user rate. An iterative algorithm based on SCA was proposed to determine the optimal solution of the joint optimization problem.
A multi-UAV cooperative mechanism was proposed in~\cite{Miao2021} to maximize the EE by the joint optimization of the UAV trajectory and transmission power. An iterative algorithm based on BCD, SCA, and Dinkelbach method was employed to determine the joint EE maximization strategy.
The work in~\cite{Sun2021} considered the the problem of minimizing of the energy consumption in a UAV-assisted relays. To achieve minimal energy consumption in the UAV, the transmit power, UAV trajectory, and time slots were jointly optimized using a iterative algorithm that is based on BCD and SCA techniques.
\paragraph{Machine Learning Approaches}
The work in~\cite{Liu2020} considered the design of an energy efficient distributed navigation system for a group of UAV-BSs for the provision of long-term coverage to ground users. They problem was first modeled as a distributed multi-agent control problem, then a deep RL~(DRL) framework was developed to control each UAV while minimizing the total energy consumption of the UAVs and ensuring minimum coverage requirements and geographical fairness are achieved.
An energy management strategy for solar powered UAV-BSs based on deep $Q$-networks was proposed in~\cite{Cong2021}, wherein the total energy consumption of the UAV-BSs as well as the 3D flight trajectory were jointly optimized to enhance their communication capacity.
The authors in~\cite{Ding2020} studied the problem of 3D trajectory design and frequency allocation while considering the energy consumption of the UAV-BSs and the fairness of user coverage. In this regard, the energy consumption of the UAV-BS was first modeled as a function of the 3D movement of the UAV. After that, a DRL algorithm based on deep deterministic policy gradient~(DDPG) was proposed to regulate the speed and direction of the UAV-BS as well as the frequency band allocation to maximize the EE, enabling the UAV-BS to arrive at its destination before the battery energy depletes and achieve fair user coverage. 

The problem of trajectory planning for multi-UAV system was considered in~\cite{Zhao2021} to achieve energy efficient user coverage. The multi-UAV trajectory planning problem was first modeled as two multi-agent cooperative games, followed by developing a decentralized cooperative RL framework based on $Q$-learning to find the equilibrium of the games. Using the proposed framework, the UAV-BSs can choose their optimal trajectory and the recharging schedule that would result in minimum energy consumption in the network.
An ML framework for joint trajectory design and power control to maximize user sum rate in a multi-UAV wireless network was proposed in~\cite{Xiao2019}. The solution procedure involves three stages. First, a multi-agent $Q$-learning algorithm was developed to find the optimal UAV positions using the initial positions of users. Second, exploiting real user mobility data set obtained from Twitter, an algorithm was developed to estimate future user locations based on echo state network. Finally, another multi-agent $Q$-learning algorithm was developed to forecast UAV positions in each time slot based on user mobility.

The authors in~\cite{Fan2020} proposed an intelligent and energy efficient framework to learn the optimal multi-UAV trajectory planning and data offloading strategy to ensure information freshness in delay sensitive applications. A multi-level DRL algorithm was developed to learn the optimal collaborative UAV control policy that would significantly minimize the energy consumption and ensure global information freshness in a dynamically changing environment.
The authors in~\cite{Abeywickrama2020} considered the problem of optimal path planning for a single UAV-BS as well as a fleet of UAV-BSs to enhance user coverage while respecting energy constraints and minimizing collisions between UAV-BSs. An RL based approach was proposed for the single UAV-BS case while a DRL based approach was proposed for the multiple UAV-BSs scenario. 

The authors in~\cite{Jianguo2021} investigated the problem of energy optimization of UAV-BSs for data collection application via a trajectory optimization technique. A $Q$-learning based algorithm was proposed to determine the optimal trajectory selection strategy that will minimize the energy consumption of the UAV while achieving coverage targets.
An ML-based framework was introduced in~\cite{Deng2019} to jointly optimize the multi-cast and trajectory for EE enhancement in UAV-based multi-cast networks. A combination of support vector regression~(SVR) and $k$-means algorithm were first proposed to determine the optimal multicast grouping, after which a centroid adjustable travelling salesman-based algorithm was proposed to determine the optimal trajectory to maximize the energy saving performance of the UAV.
A joint trajectory and power optimization scheme based on DRL was proposed in~\cite{Yuling2021} to maximize both the EE and throughput in a UAV-based communication system. The proposed scheme employs DDPG to solve the optimization problem in order to achieve the desired objective.
The CA and ML approaches that are applied for UAV-BS path planning and trajectory design in order to minimize its energy consumption are summarized in Table~\ref{tab:UAV_traj}.

\begin{table}[h!]
\centering
\caption{Summary of energy optimization techniques based on path planning and trajectory design}
\label{tab:UAV_traj}
\begin{tabular}{@{}llclc@{}}
\toprule
\multirow{2}{*}{\textbf{Paper}} & \multirow{2}{*}{\textbf{Year}} & \multicolumn{2}{l}{\textbf{Category}}                  & \multirow{2}{*}{\textbf{\begin{tabular}[c]{@{}c@{}}Specific\\ Algorithm\end{tabular}}} \\ \cmidrule(lr){3-4}
                                &                                & \textbf{CA}          & \multicolumn{1}{c}{\textbf{ML}} &                                                                                        \\ \midrule
\cite{Song2020}                 & 2020                           & \checkmark           &                                 & Heuristic                                                                              \\ \midrule
\cite{Zeng2017}                 & 2017                           & \checkmark           &                                 & Heuristic                                                                              \\ \midrule
\cite{Zeng2019}                 & 2019                           & \checkmark           &                                 & Heuristic                                                                              \\ \midrule
\cite{Ahmed2020}                & 2020                           & \checkmark           &                                 & Heuristic                                                                              \\ \midrule
\cite{Tianyu2020}               & 2020                           & \checkmark           &                                 & Heuristic                                                                              \\ \midrule
\cite{Tran2020}                 & 2020                           & \checkmark           &                                 & Heuristic                                                                              \\ \midrule
\cite{Jing2021}                 & 2021                           & \checkmark           &                                 & Heuristic                                                                              \\ \midrule
\cite{Zhang2018}                & 2018                           & \checkmark           &                                 & Heuristic                                                                              \\ \midrule
\cite{Sambo2019}                & 2019                           & \checkmark           &                                 & GA                                                                              \\ \midrule
\cite{Yang2018}                 & 2018                           & \checkmark           &                                 & Heuristic                                                                              \\ \midrule
\cite{Hua2019}                  & 2019                           & \checkmark           &                                 & Heuristic                                                                              \\ \midrule
\cite{Shivgan2020}              & 2020                           & \checkmark           &                                 & GA                                                                              \\ \midrule
\cite{Weiwei2021}               & 2021                           & \checkmark           &                                 & Heuristic                                                                              \\ \midrule
\cite{Jianping2021}             & 2021                           & \checkmark           &                                 & Heuristic                                                                              \\ \midrule
\cite{Jiang2019}                & 2019                           & \checkmark           &                                 & Heuristic                                                                              \\ \midrule
\cite{Haichao2018}              & 2018                           & \checkmark           &                                 & Heuristic                                                                              \\ \midrule
\cite{Miao2018}                 & 2018                           & \checkmark           &                                 & Heuristic                                                                              \\ \midrule
\cite{Khamidehi2019}            & 2019                           & \checkmark           &                                 & Heuristic                                                                              \\ \midrule
\cite{Hong2020}                 & 2020                           & \checkmark           &                                 & Heuristic                                                                              \\ \midrule
\cite{Xiang2020}                & 2020                           & \checkmark           &                                 & Heuristic                                                                              \\ \midrule
\cite{Lee2020}                  & 2020                           & \checkmark           &                                 & Heuristic                                                                              \\ \midrule
\cite{Dong2019}                 & 2019                           & \checkmark           &                                 & Heuristic                                                                              \\ \midrule
\cite{Gao2021}                  & 2021                           & \checkmark           &                                 & Heuristic                                                                              \\ \midrule
\cite{Bian2020}                 & 2020                           & \checkmark           &                                 & Heuristic                                                                              \\ \midrule
\cite{Eom2020}                  & 2020                           & \checkmark           &                                 & Heuristic                                                                              \\ \midrule
\cite{Miao2021}                 & 2021                           & \checkmark           &                                 & Heuristic                                                                              \\ \midrule
\cite{Sun2021}                  & 2021                           & \checkmark           &                                 & Heuristic                                                                              \\ \midrule
\cite{Liu2020}                  & 2020                           &                      & \multicolumn{1}{c}{\checkmark}  & RL                                                                                     \\ \midrule
\cite{Cong2021}                 & 2021                           & \multicolumn{1}{l}{} & \multicolumn{1}{c}{\checkmark}  & DQN                                                                                    \\ \midrule
\cite{Ding2020}                 & 2020                           & \multicolumn{1}{l}{} & \multicolumn{1}{c}{\checkmark}  & DDPG                                                                                   \\ \midrule
\cite{Zhao2021}                 & 2021                           & \multicolumn{1}{l}{} & \multicolumn{1}{c}{\checkmark}  & $Q$-learning                                                                           \\ \midrule
\cite{Xiao2019}                 & 2019                           & \multicolumn{1}{l}{} & \multicolumn{1}{c}{\checkmark}  & $Q$-learning                                                                           \\ \midrule
\cite{Fan2020}                  & 2020                           & \multicolumn{1}{l}{} & \multicolumn{1}{c}{\checkmark}  & DRL                                                                                    \\ \midrule
\cite{Abeywickrama2020}         & 2020                           & \multicolumn{1}{l}{} & \multicolumn{1}{c}{\checkmark}  & RL, DRL                                                                                \\ \midrule
\cite{Jianguo2021}              & 2021                           & \multicolumn{1}{l}{} & \multicolumn{1}{c}{\checkmark}  & $Q$-learning                                                                           \\ \midrule
\cite{Deng2019}                 & 2019                           & \multicolumn{1}{l}{} & \multicolumn{1}{c}{\checkmark}  & SVR, $k$-means                                                                         \\ \midrule
\cite{Yuling2021}               & 2021                           & \multicolumn{1}{l}{} & \multicolumn{1}{c}{\checkmark}  & DDPG                                                                                   \\ \bottomrule
\end{tabular}
\end{table}

\subsubsection{Resource Management}
In this part, we review various resource management techniques involving bandwidth and power allocation and control methods that have been developed using conventional and ML algorithms to minimize the energy consumption of UAV-BSs.

\paragraph{Conventional Approaches}
The authors in~\cite{Ruan2018} proposed an energy efficient multi-UAV deployment framework to maximize the coverage of a UAV network based on the game theory. The optimization problem was first divided into two parts: maximizing the coverage and power control with both being potential games. Then, a multi-UAV deployment algorithm based on spatial adaptive play was developed to tackle both power control and coverage maximization. 
An energy efficient power allocation scheme for UAV-enabled spatial NOMA was proposed in~\cite{Jia2021}. The EE maximization problem comprising the power consumption, signal capacity, and spatial gain was formulated, after which a heuristic algorithm was developed to determine the optimal power allocation strategy.
In~\cite{Haijun2020}, a resource allocation scheme, comprising user scheduling and power allocation in order to maximize the EE of a NOMA UAV networks under imperfect channel state information, was introduced. The proposed scheme was implemented using a heuristic algorithm for user scheduling to match the user to their sub-channels and a power allocation algorithm based on SCA was proposed to maximize the EE of the network.

The work in~\cite{Wu2021} considered the joint resource allocation and trajectory optimization in order to enhance the EE in multi-UAV based networks. An iterative heuristic algorithm based on SCA was proposed to solve the optimization problem.  
A joint power allocation and trajectory design mechanism for EE optimization in UAV-based relay systems was proposed in~\cite{Zhang2020}. The optimal power allocation strategy was determined using a heuristic algorithm based on Lagrange multiplier method while the optimal trajectory design was determined using SCA. Then, an iterative algorithm which combined both algorithms was also proposed. 
The authors in~\cite{Kaidi2021} exploited UAV for confidential information communications in the presence of eavesdroppers. To achieve this, a power splitting mechanism was proposed to enable the UAV to transmit both confidential information and artificial noise at the same time. A joint optimization problem comprising the transmit power levels, power splitting ratios, and UAV trajectory was formulated and solved using an iterative algorithm based on BCD, concave convex procedure~(CCP), and alternating direction method of multipliers~(ADMM).

The work in~\cite{Cai2019} considered maximizing the EE of a UAV-based secure communication system via the joint optimization of resource allocation and trajectory design. An iterative algorithm was developed to determine the efficient sub-optimal strategy.
In~\cite{Cai2020}, the problem of power minimization in an RIS-assisted UAV network was considered by jointly optimizing the UAV resource allocation mechanism and flight path with the help of an alternating heuristic algorithm.
The authors in~\cite{Wang2019} examined the problem of power allocation in a multi-UAV based secure network while considering both small- and large-scale channel fading. Their objective was to maximize the throughput with transmission power, total energy consumption due to transmission of each UAV and the duration of transmission as constraints. An analytical model was first developed for the throughput after which a heuristic algorithm was developed to solve the optimization problem.

A joint optimization of the UAV deployment and power allocation was carried out in~\cite{Yu2019} to maximize the area of service coverage and enhance EE. The optimization problem was decomposed into two separate problems involving the determination of the optimal transmit power control and 3D location. Using the game theory, a utility function was developed to model the cooperation between both problems, which was proven to have Nash equilibrium points. Then, a learning algorithm based on binary log-linear was proposed to find the equilibrium point in terms of coverage and EE maximization.
A resource allocation framework involving the joint optimization of time and power resources was proposed in~\cite{Masaracchia2020} to maximize the EE and throughput in NOMA based UAV network. To solve the optimization problem, a two-phase heuristic algorithm was designed: in the first phase the optimal power allocation strategy required to maximize the EE is determined while the optimal time resources to enhance the throughput of all users determined in the second phase.

In~\cite{Meng2019}, the authors examined UAV-assisted traffic offloading for EE enhancement in UAV-aided cellular networks. Their aim was to maximize the UAV-BS's EE through optimal resource allocation, trajectory design, and user scheduling. An iterative algorithm based on BCD was developed to solve the joint optimization problem.
The authors in~\cite{Xu2021} proposed a joint optimization of the power allocation and trajectory design to maximize the secrecy average rate in UAV-based secure communications. An iterative algorithm based on BCD, CCP, and ADMM was developed to solve the joint optimization problem.
An energy efficient resource allocation framework was proposed in~\cite{Mushu2020} to optimize computation offloading while ensuring the minimum energy consumption from the UAV. To achieve their target, a joint optimization of the UAV trajectory, computation task, and user transmission power was carried out, which is then solved by a heuristic algorithm based on SCA and Dinkelback.

\paragraph{Machine Learning Approaches}
The authors in~\cite{Parisotto2019} proposed an intelligent scheme based on $Q$-learning to determine the optimal power allocation and 3D position of multiple UAV-BSs during emergency situations to maximize the coverage of ground users.
In~\cite{Nan2020}, the authors addressed the problem of trajectory design and power allocation for multi-UAV networks by proposing a multi-agent DDPG algorithm to find the optimal strategy to jointly optimize the trajectory and transmission power. 
A DRL scheme for jointly optimizing the 3D-deployment and power allocation of a UAV-BS was proposed in~\cite{Meng2021}. The proposed scheme leverages DDPG and water filling to determine the optimal 3D position as well as the power allocation strategy while maximizing system throughput.
The downlink power control problem in ultra-dense UAV networks was investigated in~\cite{Li2020} for EE enhancement purposes. In this regard, mean field game~(MFG) was used to model the power control problem after which a DRL algorithm based on deep $Q$-networks was developed to determine the optimal power control policy that would maximize the EE of the UAV network.
An energy efficient resource allocation scheme for UAV-based edge computing systems that considers the power control, user association, computation resource allocation, and location planning for UAV was proposed in~\cite{Zhaohui2019}. An iterative approach based on fuzzy c-means clustering algorithm was proposed to solve the joint EE minimization problem.
Table~\ref{tab:UAV_res} summarizes the CA and ML approaches that are applied for resource management in UAV-BSs in order to minimize their energy consumption.

\begin{table}
\centering
\caption{Summary of energy optimization techniques based on resource management}
\label{tab:UAV_res}
\begin{tabular}{@{}llclc@{}}
\toprule
\multirow{2}{*}{\textbf{Paper}} & \multirow{2}{*}{\textbf{Year}} & \multicolumn{2}{l}{\textbf{Category}}                  & \multirow{2}{*}{\textbf{\begin{tabular}[c]{@{}c@{}}Specific\\ Algorithm\end{tabular}}} \\ \cmidrule(lr){3-4}
                                &                                & \textbf{CA}          & \multicolumn{1}{c}{\textbf{ML}} &                                                                                        \\ \midrule
\cite{Ruan2018}                 & 2018                           & \checkmark           &                                 & Heuristic                                                                              \\ \midrule
\cite{Jia2021}                   & 2021                           & \checkmark           &                                 & Heuristic                                                                              \\ \midrule
\cite{Haijun2020}               & 2021                           & \checkmark           &                                 & Heuristic                                                                              \\ \midrule
\cite{Wu2021}                   & 2021                           & \checkmark           &                                 & Heuristic                                                                              \\ \midrule
\cite{Zhang2020}                & 2020                           & \checkmark           &                                 & Heuristic                                                                              \\ \midrule
\cite{Kaidi2021}                & 2021                           & \checkmark           &                                 & Heuristic                                                                              \\ \midrule
\cite{Cai2019}                  & 2019                           & \checkmark           &                                 & Heuristic                                                                              \\ \midrule
\cite{Cai2020}                  & 2020                           & \checkmark           &                                 & Heuristic                                                                              \\ \midrule
\cite{Wang2019}                 & 2019                           & \checkmark           &                                 & Heuristic                                                                              \\ \midrule
\cite{Yu2019}                   & 2019                           & \checkmark           &                                 & Heuristic                                                                              \\ \midrule
\cite{Masaracchia2020}          & 2020                           & \checkmark           &                                 & Heuristic                                                                              \\ \midrule
\cite{Meng2019}                 & 2019                           & \checkmark           &                                 & Heuristic                                                                              \\ \midrule
\cite{Xu2021}                   & 2021                           & \checkmark           &                                 & Heuristic                                                                              \\ \midrule
\cite{Mushu2020}                & 2020                           & \checkmark           &                                 & Heuristic                                                                              \\ \midrule
\cite{Parisotto2019}            & 2019                           &                      & \multicolumn{1}{c}{\checkmark}  & $Q$-learning                                                                           \\ \midrule
\cite{Nan2020}                  & 2020                           & \multicolumn{1}{l}{} & \multicolumn{1}{c}{\checkmark}  & DDPG                                                                                   \\ \midrule
\cite{Meng2021}                 & 2021                           & \multicolumn{1}{l}{} & \multicolumn{1}{c}{\checkmark}  & DRL                                                                                    \\ \midrule
\cite{Li2020}                   & 2020                           & \multicolumn{1}{l}{} & \multicolumn{1}{c}{\checkmark}  & DQN                                                                                    \\ \midrule
\cite{Zhaohui2019}              & 2019                           & \multicolumn{1}{l}{} & \multicolumn{1}{c}{\checkmark}  & Fuzzy C-means                                                                          \\ \bottomrule
\end{tabular}
\end{table}

\subsubsection{Flight and Transmission Scheduling}
Here, we consider studies that involve the scheduling UAV-BSs flights as well as their sequence of transmission in order to ensure that minimum energy is consumed in the process. We also examine research works that investigate how the fight of UAVs are coordinated and scheduled or how UAVs cooperate to achieve energy efficient communication.
\paragraph{Conventional Approaches}
A flight control framework for multi-UAV systems in order to minimize the energy consumption of the UAV was proposed in~\cite{Kim2018}. The velocity control of each UAV was first modeled using a MFG theory, after which a Cucker-Smale~(CS) flocking algorithm was developed to control the velocity in order to achieve minimum energy consumption in the network.
An MDP-based framework for jointly scheduling data transmission and task computations in order to maximize the EE of a UAV-based network was proposed in~\cite{Han2021} while considering a stochastic trajectory. A probabilistic scheduling algorithm based on linear programming was proposed to solve the joint transmission and computation scheduling problem in order to minimize the total energy consumption of the network.

The authors in~\cite{Kang2020} studied the formation of swarms in multiple UAV networks to minimize the energy consumption of the UAVs. A joint optimization problem comprising transmit power, beamforming vectors, and swarm formation was formulated, after which a heuristic algorithm was developed to determine the optimal swarm formation that would result in minimum energy consumption in the UAVs.
The authors in~\cite{Xiao2020} considered the problem of EE maximization in a UAV-based relay system used for secure communication. The EE is maximized by jointly optimizing the transmission scheduling, power allocation, and UAV trajectory. The optimization problem was solved using an SCA-based iterative algorithm.
The work in~\cite{Yang2020}, studied a joint flight scheduling and resource allocation in a multi-UAV network to maximize the EE of the network. Two approaches, coordinated and uncoordinated, were proposed to solve the joint optimization problem. In the former, UAV cooperate with each other to maximize network EE via information exchange, while, in the latter, each UAV maximizes its EE separately with no information exchange. Then, heuristic algorithms were proposed to obtain the optimal solution in each approach.

The authors in~\cite{Koulali2016} investigated scheduling of beaconing periods to minimize the energy consumption for UAV-based networks during emergency scenarios. The beaconing periods of the UAVs was modeled as a non-cooperative game and then a distributed learning algorithm was proposed to determine the Nash equilibrium of the game that would result in minimum energy consumption in the network.
Similarly, the authors in~\cite{Mkiramweni2018} considered the problem of activity scheduling for UAV-BSs that are used to serve users in areas, which do not have cellular infrastructure, to maximize the EE of the UAVs. Then, a heuristic algorithm based on Nash bargaining theory~(NBG) was proposed to determine the optimal beaconing period that maximizes the EE of each UAV-BS.
In~\cite{Luan2020}, the problem of multi-task cooperation in a multi-UAV system was investigated to minimize the energy consumption of the UAV. The task assignment problem was first modeled as a coalition formation game~(CFG) and a heuristic algorithm was developed to solve the task allocation problem resulting in reduced energy consumption in the network. 

\paragraph{Machine Learning Approaches}
One of the ways of achieving energy efficient UAV operation, particularly for multiple-UAV networks, is the proper scheduling of their mobility as well as battery recharging cycle. In this regard, the authors in~\cite{Qi2020} proposed an energy efficient and fair 3D UAV-BS framework for scheduling the mobility of UAV-BSs as well as timely recharging of the UAV battery. A DRL algorithm based on DDPG was employed to solve the joint optimization of both the mobility and charging cycle of the UAV-BSs, leading to maximization of EE and achievement of fair user coverage.
The work in~\cite{Yuan2020} and~\cite{Yuan2021} considered the problem of user scheduling in UAV-based communication networks to minimize the energy consumption of the UAVs. An offline method, based on branch and bound method, was proposed to solve the problem, however, this approach involves a huge computational overhead. As a result, the problem was modeled as an MDP and an online DRL actor-critic algorithm with less computational complexity was developed to solve the user scheduling problem.
The CA and ML approaches that are applied for UAV-BSs flight and transmission scheduling in order to minimize their energy consumption are summarized in Table~\ref{tab:UAV_FTS}.

\begin{table}          
\centering
\caption{Summary of energy optimization techniques based on flight and transmission scheduling}
\label{tab:UAV_FTS}
\begin{tabular}{@{}llclc@{}}
\toprule
\multirow{2}{*}{\textbf{Paper}} & \multirow{2}{*}{\textbf{Year}} & \multicolumn{2}{l}{\textbf{Category}}                  & \multirow{2}{*}{\textbf{\begin{tabular}[c]{@{}c@{}}Specific\\ Algorithm\end{tabular}}} \\ \cmidrule(lr){3-4}
                                &                                & \textbf{CA}          & \multicolumn{1}{c}{\textbf{ML}} &                                                                                        \\ \midrule
\cite{Kim2018}                  & 2018                           & \checkmark           &                                 & Heuristic                                                                              \\ \midrule
\cite{Han2021}                  & 2021                           & \checkmark           &                                 & Heuristic                                                                              \\ \midrule
\cite{Kang2020}                 & 2020                           & \checkmark           &                                 & Heuristic                                                                              \\ \midrule
\cite{Xiao2020}                 & 2020                           & \checkmark           &                                 & Heuristic                                                                              \\ \midrule
\cite{Yang2020}                 & 2020                           & \checkmark           &                                 & Heuristic                                                                              \\ \midrule
\cite{Koulali2016}              & 2016                           & \checkmark           &                                 & Heuristic                                                                              \\ \midrule
\cite{Mkiramweni2018}           & 2018                           & \checkmark           &                                 & Heuristic                                                                              \\ \midrule
\cite{Luan2020}                 & 2020                           & \checkmark           &                                 & Heuristic                                                                              \\ \midrule
\cite{Qi2020}                   & 2020                           &                      & \multicolumn{1}{c}{\checkmark}  & DDPG                                                                                   \\ \midrule
\cite{Yuan2020, Yuan2021}       & 2020, 2021                     & \multicolumn{1}{l}{} & \multicolumn{1}{c}{\checkmark}  & Actor-critic                                                                           \\ \bottomrule
\end{tabular}
\end{table}

\subsubsection{Landing Spot Concept}
Most of the energy optimization techniques considered in the preceding paragraphs require the UAV to be in constant flight or hovering position in order to serve user requests. This greatly limits the amount of energy savings that can be obtained as the UAV consumes a significant amount of energy during flight or hovering~\cite{Fotouhi2019, mozaffari2019tutorial}. Hence, alternative deployment approaches need to be developed which can reduce the flying or hovering time of the UAV-BSs in order to further enhance the energy saving obtained from these energy optimization techniques. To address this issue, the concept of landing spot was introduced in~\cite{GangulaL1} such that the UAVs can land on some designated locations such as roof top of tall buildings, lamp post or some specially designed platforms which can also be equipped with charging pods, rather than having to hover continuously to serve user request and expend so much energy, which is a major challenge for battery limited UAVs. The few studies that consider the landing spot concept are discussed in the following paragraph. However, due to the very limited studies on this concept, the established taxonomy of categorizing the studies based on conventional and ML approaches is not followed here.

The authors in~\cite{Petrov2020} performed a capacity comparison between landed and hovering UAV with the aim of determining which approach will be suitable for adoption. Their finding reveals that the choice of a suitable deployment option depends on certain factors including the number of UAVs deployed, the distance between the charging stations and service area, and capacity of the UAV battery.
The work in~\cite{BayerleinL2} proposed a deep $Q$-learning approach for optimizing UAV trajectories using the land spot where the UAVs do not have to continuously fly along the trajectory but can land at some locations along its path in order to minimize energy consumption  while meeting user demands.
More research works need to be done in this direction to determine the optimal locations where UAVs can land along their trajectory as well as the optimal separation distances between the UAV optimal hovering positions and the suitable landing spots in order to improve the amount of energy savings while respecting the QoS constraints of the users.

\subsection{Energy Optimization of UAV-Assisted Cellular Networks}
In the previous subsection, we reviewed the cases where only the UAV power consumption was considered irrespective of whether the UAV was deployed as a standalone network or UAV-assisted cellular network. However, in this subsection, we consider various techniques, based on conventional and ML algorithms, that have been employed for the optimization of the total energy consumption of UAV-assisted cellular networks where the energy consumption of both terrestrial BSs and UAV-BSs are jointly optimized. Specifically, we focus on the literature regarding energy optimization in UAV-assisted heterogeneous networks~(HetNets).
\paragraph{Conventional Approaches}
The authors in~\cite{Chakareski2019} considered EE in a UAV-assisted mmWave HetNet comprising macro BSs~(MBSs), mmWave small BS~(SBSs) and UAVs. A joint optimization of both the sub-carrier and power allocation was carried out in order to maximize the EE of the network while considering minimum QoS and maximum transmit power as constraints. Two heuristic algorithms were developed to maximize the EE of the MBS and minimize the power consumption of the UAVs while satisfying minimum QoS requirements.
In~\cite{chang2021energy}, the authors investigated the use of UAVs for maximizing the EE in a UAV-assisted small cell~(SC) network by enabling the switch off of low EE BSs. A low complexity algorithm was developed to determine the optimal positioning and sleeping strategy that would maximize the EE of the network.
Similarly, the work in~\cite{alsharoa2017energy} considered the problem of energy optimization in UAV-assisted HetNets comprising MBSs, micro cells that can be switched off/on, and UAV SCs that are powered by solar energy. The optimization objective was to determine the optimal position of the UAVs as well as the off/on status of the micro cells that result in the minimum energy consumption in the network while respecting QoS constraints.
    
Energy and hover time optimization in a three-tier UAV-based HetNet comprising UAVs, MBS, and SBSs was considered in~\cite{Muntaha2021} to enhance the EE of the network. A two-layered optimization framework was developed where the power consumption of each tier was first optimized in order to maximize the EE of the network while respecting QoS constraints, followed by the hover time optimization of the UAVs. A heuristic algorithm based on Lagrange multiplier and subgradient approach was proposed to maximize the EE of the network.
The authors in~\cite{Manzoor2019} and~\cite{Manzoor2021} proposed a weighted power allocation scheme for minimizing the total energy consumption of both UAVs and terrestrial BSs of a UAV-assisted HetNet while satisfying QoS constraints of mobile users. A heuristic algorithm was developed to determine the optimal power allocation strategy that would result in minimal energy consumption in the network

The authors in~\cite{Pliatsios2020} considered the optimal placement of UAV mounted remote radio heads~(D-RRH) in a CRAN to minimize the transmission power while ensuring that a maximum number of users are covered. The D-RRH placement problem was first decomposed into two: vertical and horizontal placement problem. The former was solved using Weiszfeld algorithm while the latter was solved analytically.
Energy efficient UAV deployment in UAV-assisted C-RAN with flexible functional split selection was investigated in~\cite{Muntaha2021}. To achieve the minimal energy consumption, both the horizontal and vertical locations as well as the coverage radius of each UAV were optimized using mathematical analysis. In addition, a closed form expression to determine the upper and lower bound of the number of UAVs required to achieve optimal energy consumption was derived.
Table~\ref{tab:UAV_UACN} summarizes the CA and ML approaches that are applied for energy optimization in UAV-assisted cellular networks.

\begin{table}
\centering
\caption{Summary of energy optimization in UAV-assisted cellular networks}
\label{tab:UAV_UACN}
\begin{tabular}{@{}llclc@{}}
\toprule
\multirow{2}{*}{\textbf{Paper}} & \multirow{2}{*}{\textbf{Year}} & \multicolumn{2}{l}{\textbf{Category}}         & \multirow{2}{*}{\textbf{\begin{tabular}[c]{@{}c@{}}Specific\\ Algorithm\end{tabular}}} \\ \cmidrule(lr){3-4}
                                &                                & \textbf{CA} & \multicolumn{1}{c}{\textbf{ML}} &                                                                                        \\ \midrule
\cite{Chakareski2019}           & 2019                           & \checkmark  &                                 & Heuristic                                                                              \\ \midrule
\cite{chang2021energy}                & 2021                           & \checkmark  &                                 & Heuristic                                                                              \\ \midrule
\cite{alsharoa2017energy}             & 2017                           & \checkmark  &                                 & Heuristic                                                                              \\ \midrule
\cite{Muntaha2021}              & 2021                           & \checkmark  &                                 & Heuristic                                                                              \\ \midrule
\cite{Manzoor2019, Manzoor2021} & 2019,2021                      & \checkmark  &                                 & Heuristic                                                                              \\ \midrule
\cite{Pliatsios2020}            & 2020                           & \checkmark  &                                 & Heuristic                                                                              \\ \midrule
\cite{Wang2018}                 & 2018                           & \checkmark  &                                 & Heuristic                                                                              \\ \midrule
\end{tabular}
\end{table}

\section{Enabling technologies for energy efficiency in UAV-based cellular Networks}\label{sec:enablers}
In this section, we present some of the technologies that empower UAVs to achieve high EE when used in cellular networks.
Any technique that enables the UAV to process less data, focus its energy more or extract energy from the ambient environment will prolong its operational lifetime and is therefore considered energy-efficient. Below we discuss such technologies and show how they enable UAVs to achieve energy savings.

\subsection{RIS}
RISs are meta-surfaces that use electronic circuits (usually varactor diodes or other micro-electromechanical systems) to manipulate impinging electromagnetic signals in a desired sort of way.
By intelligently manipulating each reflecting surface, RISs can change the amplitude, phase or frequency of each incoming signal and redirect  them to a desired location without using power amplifiers. 
In an RIS, each meta-surface is capable of independently steering an impinging electromagnetic ray to a desired angle of reflection, and manipulating the phase of the reflected ray so that it aligns with the angle and phase of other rays from other meta-surfaces in the RIS~\cite{RISBasar2019}, thereby ensuring that all the reflected rays are received with the same angle and phase (these are usually designed to align with the angle and phase of the line-of-sight signal (LOS)) at the receiver.
By acting in such a manner as a low complexity (re-)~transmitter, RISs can forward signals to a receiver and have been shown to largely outperform other amplify-and-forward-based relaying techniques in terms of received power (mostly due to their large geometry of the RIS antennas)~\cite{RISBasar2019} as well as EE in wireless communication systems~\cite{RISEnergyEfficiencyHuang2019}, since they do not require active elements.
They are mostly passive and do not require external power supply. However, they do not amplify the RF signals impinging on their surface~\cite{RISBasar2019}, which is also an advantage since they do not amplify the noise.

RISs can be combined with UAVs to improve system throughput~\cite{RISUAVCommunicationLi2020}, network coverage and reliability~\cite{RISDualHopUAVCommunicationYang2020, URLLC_RIS_UAVRanjha2021} or EE~\cite{MLTrajectoryBeamformingDesignRISUAVLiu2021, EnergyOptimizationRISUAVMichailidis2021energy, TrajectoryBeamformingDesignRISUAVLong2020, RISUplinkDiamanti}.
By acting as relays for signals without the need for additional external power supply, RISs achieve high EE when combined with UAVs in cellular networks.
RISs are deployed between the UAV and ground-based user nodes. As a result, the UAV can transmit at a much lower power to reach the RISs~\cite{RISBasar2019}, leading to significant power savings.
In addition, since received power is proportional to the effective area of the receiving antenna, RISs can capture more of the transmitted RF signal compared to user devices that have much smaller antennas. Again, this enables the UAV to lower its transmit power due to the efficiency of the RISs, which helps the UAV to conserve its power and prolong its operational lifetime.

UAVs often need to move to maintain LOS connection with mobile ground terminals.
With RIS, the UAV simply beams its signal on the reflecting surface, whose reflecting angle and phase are then manipulated to focus the signal in the direction of the mobile users.
This has been shown to significantly lower UAV energy consumption since up to 95\% of the energy consumed by rotary-wing UAVs is spent on flying~\cite{MLTrajectoryBeamformingDesignRISUAVLiu2021}.
It was also shown in~\cite{UAVRISVehicularNetworksAgrawal2021} that even in the presence of a direct LOS signal between the UAV and internet-connected vehicles, RIS significantly improves the bit error rate, coverage probability, and throughput. These metrics imply not only less transmission power from the UAV but also fewer retransmissions.

\subsection{Mobile edge computing (MEC)/Cloud}
MEC can prolong the operational time of UAVs by offloading some computing tasks from the UAV. The concept of allowing UAVS to offload computational tasks to MEC servers was studied in~\cite{UAVMECSurveillanceMotlagh2017}, where it was shown to improve both EE and processing time for assigned tasks.
Energy saving for UAVs through computational tasks offloading to MECs was extended to the case of multiple UAVs in~\cite{MultipleUAVMECYang2019}, where the authors proposed a system to minimize both the computation, communication, and mechanical (due to flight or hovering) energy consumption of UAVs subject to latency and coverage constraints.
The technical challenges that must be overcome to actualize UAV-based MEC include UAV trajectory/position planning~\cite{MECUAVTrajectoryRA_Li2020, UAVMECTaskPredictionLSTMWu2020, BitAllocationTrajectoryLi2019}, allocation of computational loads~\cite{BitAllocationTrajectoryLi2019} in the case of multiple UAV/MEC servers, user transmit power/interference management~\cite{UAVMECNOMAZhang2020}, security against eavesdropping and jamming~\cite{SecurityUAVMECBai2019, UAVMECSecurityRLChen2021}, etc. while minimizing the energy consumption of the UAV (and in some cases that of the users~\cite{BitAllocationTrajectoryLi2019}) so as to prolong its operational life.
The authors in~\cite{WorkflowSchedulingUAV_MECDu2019} proposed a time division multiple access (TDMA)-based workflow scheduling model to improve UAV EE while offloading computations to MEC servers by optimizing user association, computing resource allocation, hovering duration, and service sequences of users.
Offloading computing tasks to UAV MECs and switching off computational elements was shown to lower energy consumption and job loss in~\cite{5GSlicingMECUAVsRLFaraci2020}, wherein the authors implemented an RL model to manage a framework responsible for switching computing elements in order to lower power consumption, delay, and job loss.

The common architecture in UAV-based MEC systems involves tethering the MEC server to the UAV. This way, the UAV provides coverage and connectivity to the users on ground, while all computational tasks are offloaded to the MEC server. This scheme was implemented in~\cite{UAVMECEnergyHarvestZhou2018}, where the UAV also provided wireless energy harvesting services to the ground users in addition to communication and computation coverage.
Similarly, it was shown in~\cite{MobileCloudletJeong2018} that, instead of serving just as a flying BS or mobile relay for ground-based users, the UAV can function as a mobile cloudlet to run compute-intensive applications for the users. This is achieved by
offloading such tasks to a cloudlet attached to the UAV, thus minimizing the overall mobile energy consumption without reducing the QoS requirements of the target applications. 

Some of the articles highlighted above have adopted state-of-the-art ML algorithms to improve EE for UAV-enabled MEC networks. A few examples include the work in~\cite{UAVMECTaskPredictionLSTMWu2020} where the long short term memory (LSTM) algorithm was employed to predict computational tasks from the users and thereby direct MEC computational resources accordingly, and in~\cite{UAVMECSecurityRLChen2021} where RL was employed to plan an allocation strategy for the computational tasks whereas a transfer learning algorithm builds on the allocation strategy to achieve faster computation. Similarly, a cooperative UAV MEC system was implemented in~\cite{5GSlicingMECUAVsRLFaraci2020}, where a UAV can offload computation to nearby UAVs to conserve energy and reduce latency. A RL algorithm was employed to
assist a system controller in allocating the tasks to be done to the available UAVs to improve target metrics of EE, latency, and job loss.

\subsection{Network Slicing/Network Function Virtualization} 
Network function virtualization (NFV) is a networking paradigm whereby network functions (such as address translation, firewall services, storage, routing, deep packet inspection, etc.) are decoupled from the underlying hardware and instead realized as software (thus the term, \textit{network softwarization}) that can be run on standard commercial off-the-shelf servers/equipment~\cite{NFVSurveyYi2018}, and accessed via standardized interfaces.
NFV helps to improve the flexibility, resilience, EE and management of UAV-based networks. The softwarisation of network functions leads to less processing at UAV-BSs, which significantly improves the energy usage of such BSs.

In such software-defined network scenarios, the automation of network management functions will free the UAV from performing those functions, so that its on-board energy can be dedicated to performing other tasks~\cite{NFVSchedulingTipantuna2019}.
For instance, the authors in~\cite{DynamicSpectrumManagementNFVXu2021} introduced an NFV scheme where spectrum processing, allocation, and management are dynamically implemented as virtualized functions to serve UAV-BSs. Cognitive radio services are also virtualized to enable the UAV to provide both cellular, Wi-Fi, and WIMAX services according to the application requirements for the coverage area. Offloading storage and processing functions to the ground control stations gives the UAVs higher endurance.
NFV not only allows UAVs to save energy but also to be rapidly adapted for different purposes based on application demands.
This tack was taken in~\cite{ConfigurableUAVsNogales2018} where small UAVs were programmed to be deployed by a network operator to offer any chosen service such as flight control, voice over IP telephony, routing, etc. The authors demonstrated a prototype of using such a system to provide IP telephony services.

In implementation, a NFV infrastructure is used to host all the network shared hardware and virtual resources. In fact, by allowing a diverse set of network services to be run remotely on such shared infrastructure, NFV inherently enhances resource utilization~\cite{SoftwarizationUAVsSami2020}, including on-board energy resources in UAVs.

\subsection{Cooperative Communications}
Cooperative communication helps to lower the energy consumption of UAVs by employing other devices within the network or other UAVs to assist in data transmission, instead of using the UAV to route all network traffic.
For instance, UAVs serving as relays (for one another or between an eNodeB and ground-based terminals) can help to improve network coverage~\cite{UAVRelaysPlacementChen2018}, reliability~\cite{UAVReliabilityRelayAzari2018} or bypass unfavourable channels. That way, data is always relayed through favourable channels between communicating pairs~\cite{UAVCooperativeTimeSharingRelayYin2020}, leading to both proximity gain (higher throughput) and significant energy savings. Proximity gains result from the fact that UAVs are deployed closer to user terminals than MBSs, especially when are deployed to extend the coverage of the MBS.
The energy savings in this case are due to the proximity/nearness of the UAVs to the terminals, allowing them to transmit with lower powers and reducing re-transmissions (the probability of error is low when the communicating terminals are close), thereby achieving significant energy savings~\cite{CommunicationDistanceUAVRelayThammawichai2018}. In addition, cooperative communication leads to spatial diversity gains since the relays employed in such systems receive and process the signal from different locations than the original transmitter. Thus, cooperative communication improves the system reliability by ensuring that the signal is delivered to the destination even if a channel is in deep fade.
It has been shown that in some implementations, cooperation achieves full diversity~\cite{CooperativeDiversityLaneman2004}, which can significantly suppress the interference arising from multiple terminals~\cite{CooperativeCommsLetaief2009}, leading to fewer re-transmissions that indirectly leads to higher energy savings. An example is shown in~\cite{UAVWayfindingHo2013} where the authors demonstrated that using cluster heads as cooperative relays lowers the energy consumption of a UAV.

RISs can also be viewed as a cooperative communication mechanism when combined with UAV networks to improve cellular coverage and throughput. They save energy by relaying transmitted data from the source to the destination. RISs not only reduce the transmit powers needed by UAVs to reach ground terminals but also improve coverage and suppress channel fading. This has been shown to improve the system EE~\cite{RISUAVResourceAllocationNguyen2021} in general and to conserve the energy of the UAV in particular.

\subsection{Energy Harvesting Technologies}
Energy harvesting has become a hot topic among the scientific research community. As more and more power-hungry internet-connected devices are deployed in the IoT era, finding enough power sources to feed them is becoming a technical as well as an environmental problem. In addition, since most wireless devices are mobile, using stationary cables to recharge their batteries limits their mobility and therefore, their utility~\cite{WIPTandEHStatusProspectsHuang2019}.
However, there are numerous energy sources in the ambient environment that can be harnessed to provide power to these devices to keep them operational for very long periods of time (if carefully designed, some devices can harvest enough ambient power to potentially run perpetually.)

Popular energy harvesting schemes include harvesting energy from the electromagnetic fields, solar energy from the sun using photovoltaic cells, mechanical energy due to vibration or motion of the UAV, thermal energy, optical energy using laser beams~\cite{WIPTandEHStatusProspectsHuang2019}, etc. In electromagnetic waves-based energy harvesting (e.g. RF energy harvesting), the transmitter coverts some of the direct current (DC) from its battery into the RF energy to be transmitted. At the receiver, the received RF energy is converted back to DC and amplified using matching and rectifier circuits~\cite{RFSOlarHarvestingQuyen2020}, then used to charge the on-board battery or directly used to power decoding or other communications operation. The efficiency of these RF-DC conversions significantly affects the amount of energy harvested.

In~\cite{UAVCooperativeTimeSharingRelayYin2020}, a UAV scavenges RF energy from a transmitting source (such as an eNodeB) to partially power the energy used in communication to enable it to serve as a relay to forward data between the source and user terminals on the ground.
The energy harvesting is based on the time-splitting technique in which the UAV harvests energy in one time slot and performs communication in another slot.
A similar energy harvesting technique for UAV-supported communication was explored in~\cite{UAVCooperativePowerSplittingRelayYin2019} wherein power-splitting technique was used instead of the time-splitting scheme adopted in~\cite{UAVCooperativeTimeSharingRelayYin2020}. In power-splitting, the energy harvesting unit (UAV in this case) has separate hardware for handling communication and energy harvesting, so that both can be performed simultaneously.

In addition, the authors in~\cite{TimePowerAllocationUAVHarvestLiu2020} derived an optimal strategy for allocating charging time and power to UAVs to wirelessly charge them during operation.
The charging power is supplied by the BS i.e., it is scavenged from the RF energy typically used for communication.
A different tack was followed in~\cite{AerielEnergySharingUAVsLakew2021} where the authors proposed an energy sharing scheme whereby high-capacity UAVs harvest solar energy and share it with smaller-capacity UAVs.
In some implementations of energy harvesting to support UAV operations, the authors assume that the harvested energy is sufficient to support the communication functions of the UAV, whereas in others, harvested energy merely serve to support power provided by the UAV on-bard battery.
In~\cite{UAVRelayingHarvestYang2018}, Yang et al. studied the outage problem of an UAV that harvests energy from a ground-based BS and uses the harvested energy to relay data for user terminals on the ground.

\section{Challenges and Open Research Problems}\label{sec:challenges}
In this section, we present the most common challenges and open research issues that are limiting the widespread adoption of UAVs in cellular networks. The list presented is not exhaustive but is meant to guide the reader on some of the most pressing concerns that need to be addressed to fully exploit the advantages of UAVs as an aid to cellular networks.
The reader is referred to the work in~\cite{DesignChallengesMultiUAVShakeri2109}, which for a more detailed look at UAV network design challenges.

\subsection{Security Challenges}
Security is a big challenge in UAV-aided communication. Due to the small form factor and capability of most UAVs, they are prone to both cyber and physical security threats, such as hijacking.
Moreover, due to the broadcast nature of UAV communications, they are subject to eavesdropping, packet snooping, jamming, spoofing, denial of service attacks, and cyber attacks~\cite{BlockchainUAVsMehta2020, Fotouhi2019}. 
The security of UAV-to-UAV communication against multiple eavesdroppers was investigated in~\cite{SecurityUAV2UAVYe2019} where the authors derived expressions for the statistical SNR as well as for the secrecy outage probability.
In extreme cases, interceptors could potentially spoof the control signals and use them to gain control of the UAVs~\cite{SecureUAVLi2019, Fotouhi2019}, or jam such signals to prevent them from reaching the ground control station.
Thus, authentication of users and operators is an issue in UAV-based networks~\cite{ANNCellularUAVChallita2019}.
UAV security issues are categorized under cyber detection, which focuses on identifying intruders within the UAV network, and cyber protection, which focuses on eliminating or reducing external threats to the UAV network~\cite{DetectingUAVCyberAttackSedjelmaci2016}. 

The authors in~\cite{UAVOSSecurityIqbal2021} also distinguished different forms of security, such as information and software security, sensors security, and communications security.
Information and software security is related to the security of the UAV operating systems (including its configuration files, mission-related data, and flight control files) as well as its collected data.
Sensor security deals with the security of the various sensors used for the real-time maneuvering of the UAV, such as accelerometers, gyroscopes, GPS, barometers, etc. If attackers gain control of such subsystems, it could lead to the malfunctioning of the UAV or cause it to send erroneous data.
Communications security is related to the security of the communications components of the UAV, including control commands, telemetry data, and transmission of the collected data.
Since these communications happen over the air and some of the packets are unencrypted (e.g., data packets from simple sensors or even GPS data), they pose cybersecurity threats to the successful operation of UAV networks.

Packet routing is another potential source of  attacks on UAV communications.
Here, malicious entities can disguise themselves as legitimate elements in the UAV network to steal, modify or drop packets.
There are three common types of routing-related attacks: wormhole, selective forwarding and sink hole attacks~\cite{SecureCommsUAVFotohi2020}.
Wormhole attacks involve a malicious entity that captures packets in one location within the network, then tunnels the packet to a third party in another location where the packets are modified and retransmitted into the network~\cite{WormholeAttacksYihChun2006}. This attack is common in ad hoc networks that use reactive routing protocol such as shortest-path-first routing, where the routing metric is based on the hop count. Malicious intermediaries pretend to be neighbour nodes of genuine network nodes, and route packets through private tunnels. Tunneling ensures that the packets arrive at the destination node with shorter distances or lower number of hops than genuinely multi-hop-routed packets, thereby misleading the receiver.
Selective forwarding attacks occur when malicious nodes behave like genuine network nodes and correctly forward packets most times, but occasionally selectively drop sensitive packets that may be critical to the functioning of the network. In UAV networks, other UAVs or edge devices can be used to perpetrate selective forwarding attacks.
Since wireless networks randomly drop packets (due to congestion or unreliable links), this type of attack is difficult to detect because it is hard to distinguish if packet drops are due to network faults or malicious attacks~\cite{SelectiveForwardingAttacksRen2016}.
In sink hole attacks, adversary nodes advertize a false hop count to the sink, tricking its neighbouring nodes into believing that they have found a shorter route to the sink and then forward their data packets via the adversary node~\cite{SinkHoleAttacksLiu2020}.
Thus, the adversary node not only gains access to packets of nodes within its radio range but also blocks these nodes from transmitting to the final receiver or genuine network sink. Again, other UAVs, relay nodes or edge devices can be used to perpetrate this attack in UAV-assisted networks. 

A potential solution for the communication security challenges in UAV-based networks is to use blockchain technology~\cite{BlockchainUAVsMehta2020}.
Proper authentication will also ensure that only vetted elements are admitted into the network.
Designing a robust intrusion and detection system will also protect UAV-aided networks from malicious attacks~\cite{UAVOSSecurityIqbal2021}.
In addition, innovative communication protocol design such as use of spread spectrum technologies, MIMO techniques or cooperative UAV features can also help to protect UAV communication from eavesdroppers. Due to their directionality, mm-wave communication can also reduce the threat of jamming and eavesdropping since the electromagnetic signal beams are focused on the intended receiver(s)~\cite{PHYSecurityUAVsSun2019}.
Data encryption is very helpful in dealing with information and data security, as is proper design of the configuration files through a process known as \textit{system hardening}~\cite{UAVOSSecurityIqbal2021}.
Wormhole attacks can be detected and defended using packet leashes~\cite{WormholeAttacksYihChun2006}, where some information is added to a packet to restrict the maximum transmission distance per link.
In addition, ANN have been proposed to create threat maps in the operating environment of UAVs, with recurrent neural networks used to predict the normal trajectory motion of the UAV and hence detect any deviations in real-time as suspected cyber attack~\cite{ANNCellularUAVChallita2019}.

\subsection{Complexity}
Increasingly, UAVs are deployed as a swarm of vehicles to achieve communication, reconnaissance, search/rescue or monitoring tasks. Due to the complexity of these tasks, multiple UAVs (also called a swarm of UAVs) are usually required to perform them more efficiently. The advantages of using multiple UAVs in such scenarios, such as cost and time efficiency, reliability/fault tolerance, flexibility/adaptability to changing requirements, and the ability to perform multiple tasks simultaneously, have been well-established~\cite{MultiUAVSurveySkorobogatov2020}.
However, due to the dynamism and uncertainty of the operating environments for such complex tasks, coordinating multiple UAVs to work together is highly challenging~\cite{MultiUAVCoordinationCommunicationsTortonesi2012}.
In addition, UAVs nowadays are increasingly subjected to more and more complicated cyber and physical threats. Thus, they have to be designed with additional software and hardware components to thwart such threats, which further increases their complexity.
The most common issues that arise in multi-UAV coordination have been documented in~\cite{Fotouhi2019, AerialSwarmRoboticsSurveyChung2018, InterferenceCoordinationUAVsShen2020, LiveFlyUAVExptChung2016, ChannelSlottingMultiUAVChen2018}. They include: 
\begin{itemize}
    \item Algorithmic planning to manage communication and task allocation,
    \item Coverage issues and equitable distribution of workload,
    \item Aeriel manipulation of the vehicles,
    \item Power management,
    \item Management of the communication infrastructure,
    \item Path planning to avoid collisions while ensuring adequate coverage without overlaps,
    \item Interference arising from other UAVs,
    \item Conflict resolution,
    \item Safety issues related to preventing the vehicles from flying into one another's buffer zones,
    \item  Safety issues related to take-off and landing (in some current implementations, a swarm of fixed-wing UAVs spent less than 20\% of the time staying simultaneously in the air to execute assigned tasks) while the bulk of the time is spent trying to coordinate the flight of the UAVs.
    \item Network congestion and channel interference~\cite{} due to multiple UAVs exchanging data to coordinate the execution of assigned tasks.
\end{itemize}
Multi-UAV systems may also require more than a single pilot to manage them, which introduces another layer of complexity to the system~\cite{MultiUAVSurveySkorobogatov2020}.

One of the main challenges facing the implementation of a swarm of UAVs is localization. To be effective as part of a fleet, a UAV needs to be aware of its position in a given map of the environment. The UAV position is either relative to a reference point or relative to other UAVs in the fleet.
As one can imagine, this is a non-trivial task that requires numerous exchanges of communication and control commands. In addition, since the positions of the UAVs are constantly changing, the map of the fleet is also constantly changing 3D. This gives rise to a dynamic 3D map of the environment, rather than a constrained static map with a reference landmark~\cite{AerialSwarmsChallengesAbdelkader2021}. Thus, achieving partial or full localization is both energy- and bandwidth-intensive.
GPS sensors are insufficient to address this problem because they provide position accuracy to only three meters~\cite{AerialSwarmsChallengesAbdelkader2021}, which is not granular enough to prevent collisions. Alternatives found in the literature include equipping the UAVs with wireless communication modules and inertial navigation systems, coupled with on-board sensor fusion to enable accurate estimation of position. As highlighted already, this incurs a heavy computational, communication (bandwidth), and energy cost, hence, innovative ways to do this more efficiently is still an open problem.

Path planning is another challenge that arises when multiple UAVs work together to achieve a common objective (this is related to the localization problem). As the number of vehicles in the swarm/fleet grows, it becomes more difficult to plan the trajectory of each UAV from the starting points to the goal points in order to traverse the minimum path (so as to save energy) and avoid collisions with obstacles or other vehicles in the swarm.
In addition, path planning must be executed such that the UAVs maintain connectivity with one another and with the ground control station while performing assigned tasks.
Path planning involves motion planning (to control the path length and turning angles), trajectory planning (involving the speed and kinematics of the vehicle) and navigation (involving localization and obstacle/collision avoidance)~\cite{PathPlanningTechniquesAggarwal2020}.
For UAV swarms used as aeriel BSs, path planning in such cases requires high-rate exchange of positional sensor data, which involves multi-dimensional channel characterisation, tracking and communication, interference management, transmit power allocation, resource block assignment, etc.~\cite{CooperativeVehicularNetsZhou2015}.
Path planning for other UAV applications such as monitoring, or target tracking are complicated by issues such as target location and identification which arise due to the high mobility of the UAVs.
Some of the techniques used in path planning can be categorized under representative, cooperative, and non-cooperative techniques~\cite{PathPlanningTechniquesAggarwal2020}. Increasingly, ML approaches are employed to address path planning problems~\cite{PathPlanningZear2020}.
To save on energy, time, computational and communications costs, path planning is a complex task whose complexity grows rapidly as the number of UAVs in the swarm increases.

Other key challenges that arise due to the complexity of multi-UAV systems include security, equitable allocation of workload and coverage area.
The higher the number of UAVs in a swarm, the larger the attack surface~\cite{ThreatModelingUAVsAlmulhem2020}. Therefore, modelling potential threats and designing a robust security framework to thwart cyber attacks is a highly complex and challenging task.

\subsection{Data Availability}
ML algorithms live on data; the more data we feed the algorithms, the smarter they tend to become~\cite{MLDataVisualizationLi2018}.
To make strategic and intelligent decisions, data is typically required from different sources; these data are integrated and transformed before they are used to make decisions or predictions~\cite{DataIntensiveBiasPark2021}. This process is the so-called data-driven or evidence-based decision-making~\cite{MLTrendseJordan2015}.
In many cases, large data sets are required both to train and test the ML algorithms as well as to evaluate their efficacy in making predictions.
In addition, both labeled and unlabelled data are required to compare the efficacy of different ML algorithms~\cite{MLWirelessNetworksSun2019}, and guide in selecting the best technology tool for a given situation.
However, data is not always available or is available in insufficient quantity or quality. 
Due to the criticality of data to the success of businesses, companies are usually reluctant to share data obtained from UAV deployment trials.
There are also cases where it is very difficult to obtain data, especially when a process or technique is still in infancy.
Some of the most useful data sets are not publicly available.
To make predictions which are generalizable, there needs to be sufficient supply of heterogeneous data repositories that are accessible many researchers. Failing this, each institution might end up developing individual analytical pipelines that may fail under new circumstances, such as a different operating environment.

Open access data repositories are a key enabler to unlocking insights into addressing the challenges of UAV deployment in cellular networks.
With sufficient data shared collaboratively amongst stakeholders in the UAV research and development industry, robust and reproducible ML solutions can be developed for UAV applications in cellular networks.
The ability to obtain and reuse data will enable easy collaboration among researchers and the industry, save costs, and minimize time to market for new products.
It is desirable to have an open database for UAV-aided cellular communication, similar to the modified National Institute of Standards and Technology (MNIST) database (\textit{c.f.}~\cite{MNISTDeng2012}) widely used in computer vision studies.

\subsection{Limited Energy Storage Capacity}
One of the key challenges of using UAVs to support communication networks operations is that they have limited lifetime due to the low capacity of their batteries, which limits how long they can be deployed~\cite{UAVSwappingBhola2021}.
In fact, most UAVs have an endurance of just a few hours~\cite{UAVITSChallengesMenouar2017}.
To reduce the weight of UAVs and the attendant energy drain problem, it is often necessary to use smaller batteries so as to avoid expending too much energy on flying as the energy required to fly the UAV varies with the payload size~\cite{EMFUAVChargingNguyen2020}.
However, small batteries have low storage capacities, further complicating the energy situation of the UAV.
In addition, there exists uncertainties in estimating the remaining battery charge in UAVs~\cite{BatteryHealthUAVSaha2011}, leading to conservative estimates of left-over flight time to avoid \textit{dead stick conditions} whereby the UAV runs out of battery power in-flight, which could have disastrous consequences.
Supercapacitors, on the other hand, are not ideal for UAVs due to their low energy density~\cite{simic2015investigation}. 
Improvements in battery technology are required to enhance the storage capacity of UAVs.

One of the solutions that have been proposed to address the battery capacity limitations of UAVs include wireless charging~\cite{simic2015investigation}, whereby the UAV is recharged during operation via  RF energy harvesting~\cite{EMFUAVChargingNguyen2020}, laser power beaming~\cite{LaserChargedUAVJaafar2021, LaserPowerXmissionJin2019}, use of electrical power lines~\cite{UAVPowerlinesChargingLu2017} by directly perching on current-carrying cables or harvesting electromagnetic energy generated by the cables.
Other alternatives for recharging the UAV when its battery becomes depleted have been explored in~\cite{UAVRechargingOptionsGalkin2019}, including UAV swapping (replacing those with depleted batteries with others that are fully charged)~\cite{UAVSwappingBhola2021} and battery swapping. Alternatively, the UAV could be tethered to a mains power supply~\cite{TetheredUAVSaif2021} or powered with fuel cells~\cite{boukoberine2019critical}.

\subsection{Energy Harvesting Challenges} 
It is well known that one of the biggest limitations to the widespread adoption of UAVs in cellular networking is their limited operating lifetime~\cite{PowerTimeAllocationUAVEHLiu2020}.
One of the most popular solutions for dealing with this problem is using energy harvesting technologies to enable UAVs to harvest energy from the environment to support the on-board battery.
However, this technology is still in its infancy and fraught with many natural and technical challenges. For instance, solar-based energy harvesting still depends on climatic conditions and become unavailable when the sun is not shining~\cite{WIPTandEHStatusProspectsHuang2019, lu2018wireless}, especially in the winter (which can last several months in the northern hemisphere). Solar energy harvesting solutions are based on use of photovoltaic arrays, which are only suitable for fixed-wing UAVs~\cite{boukoberine2019critical}. Similarly, the availability of RF energy depends on the density of RF devices within the area.
Most of the energy harvesting technologies in the literature have low energy transfer efficiency due to environmental (such as free space pathloss for RF energy harvesting) or device limitations (such as poor energy conversion ratio). Moreover, since UAVs are usually in motion, they keep losing LOS connection with the charging BS.
As a matter of fact, the amount of energy harvested is still very small compared to the amount of power that can be stored in on-board batteries, especially when such energy comes from RF environment~\cite{UAVEHInterferenceChen2019}. 

The output of most energy harvesting techniques is still quite low due to the poor efficiency of the energy conversion and matching circuits~\cite{RFEfficiencyEfficiencyCansiz2019}.
The amount of energy captured from RF energy, for instance, depends on the area of the antenna elements~\cite{RFSOlarHarvestingQuyen2020}.
Since most UAVs come in small form factors to limit their weight, their antennas are also small. Even where solar or other forms of energy are targeted for energy harvesting, the small form factor also implies that a limited area of the UAV is exposed to ambient energy, thereby limiting the amount of energy that can be harvested from such environments.
Moreover, the distance between the UAVs and the charging (base) station as well as whether the UAV has an LOS connection to the base also affect the amount of energy received; the farther the UAV, the lower the energy received~\cite{AerielEnergySharingUAVsLakew2021}. However, since UAVs are deployed to supplement the services provided by MBSs, it is desirable that they operate far from the BS in areas with poor coverage, which limits the power received from the macro cell.

Another limitation associated with energy harvesting is the time it takes to recharge in-flight UAVs from harvested energy due to shortcomings in the charging rates.
Finally, since energy harvesting is still in its infancy, there is a lack of a unified standard on it, which limits the widespread involvement of original equipment manufacturers (OEM) to develop wireless recharging devices.

One way to improve the amount of energy harvested from the radio waves is to use directional antennas via beamforming~\cite{AerielEnergySharingUAVsLakew2021, BeamformingEnergyTransferChoi2017}, which ensures that more of the transmitted RF energy is received at the UAV.
There are other proposed solutions aimed at optimizing the charging time of UAVs by improving the energy transfer  and conversion rates. Solutions related to the above two include optimal or near optimal trajectory planning to guarantee maximum exposure of the UAV to the RF field of the charging BS in cases where energy is harvested from the radio environment.

\subsection{Regulations}
Due to public safety, privacy, and data protection concerns, there are strict restrictions on use of UAVs by governments around the globe.
Both international and national regulations are put in place to ensure that UAVs pose minimal risks to other users of the airspace and to protect people and property on the ground~\cite{UAVRegulationsReviewStocker2017}.
For UAVs to gain wider acceptability, the public needs to trust that their deployment is in their best interests and will be used safely.
For instance, there are valid concerns arising from the knowledge that UAVs have been used for illegal activities such as surveilling or tracking people.
In addition, there are concerns over data protection since UAVs can be equipped with cameras and other sensors that can collect data from areas where they do not have authorization.
There are also serious concerns regarding public safety, as UAVs can lose control during flight and collide with other aircraft, buildings, etc. or crash into people on the ground~\cite{Fotouhi2019}.
Moreover, UAVs can be equipped with weapons and used to carry out remote attacks.

There are many regulatory barriers that need to be addressed before the full potential of UAVs as part of wireless networks can be realized.
One of the commonest issues is restrictions on areas where UAVs can operate.
Due to some of the concerns highlighted above, UAVs serving as cellular BSs are not yet approved to operate in many public areas, especially in areas with large crowds.
In addition, there are strict restrictions on how far UAVs can fly.
Even when a UAV is capable of operating autonomously, many municipalities and cities still require that there be a licensed UAV pilot present before a UAV can be deployed, which increases the operating costs of deploying such UAVs and limits their use.
The red tape in the approval of the use of UAVs has also been highlighted as a challenge mitigating against their widespread use.
Regulatory and policy issues are usually slow and far behind the advances in UAV development, which in turn curtail the research, development and deployment efforts by both the telecommunication industry and the research community~\cite{UAVRegulationsReviewStocker2017}.

Despite some of the current limitations, significant progress has been made in UAV regulations. In Europe, for example, roadmaps have been created on how to integrate UAV operations into the civilian aviation industry~\cite{UAVRegulationsReviewStocker2017}. There are similar efforts around the globe to enact regulations that will both promote the wide adoption of UAVs in civilian applications as well as ensure privacy and public safety~\cite{UAVRegulationsReviewStocker2017}.

\section{Conclusion}\label{sec:conclusion}
This survey paper covers the energy optimization techniques for UAV-assisted wireless communication networking by categorizing them in terms of the optimization algorithm employed.
On one hand, there are some well-known optimization methods, such as heuristics, game theory, etc., that have been widely used for energy optimization.
The implementation of ML for optimization, on the other hand, have been gaining momentum due to its proven capabilities.
Thus, we combined both conventional and ML algorithms in this survey paper in order to cover the literature in a comprehensive and inclusive manner.
The studies on energy optimization in UAV-assisted wireless networking were investigated thoroughly to reveal the state-of-the-art.
Some background information on both the optimization algorithms and power supply/charging mechanisms of UAVs were given in order to cover the topic in a more complete manner.
Moreover, different types of UAV deployments were also discussed to highlight how the UAV-assisted communication networks can be divergent, increasing the level of challenge in optimization.
As one of the most novel parts of this survey, emerging technologies, such as RIS and landing spot optimization, were presented to capture the latest advancements in the literature.
The survey was concluded by the identification of challenges and possible research directions.
This will help focus the research efforts into these areas, thus making the UAV-assisted wireless networking a complete and mature concept.
This, in turn, would result in a feasible and applicable concept for wireless communication networks, which has the potential to mitigate the capacity scarcity issue.
\bibliographystyle{IEEEtran}
\bibliography{ref}

\begin{IEEEbiographynophoto}{Attai Ibrahim Abubakar} obtained his B.Eng. degree in Electrical and Electronics Engineering from Joseph Sarwuan Tarka University (formerly Federal University of Agriculture), Makurdi, Nigeria, in 2011 where he graduated topmost rank, and M.Sc. degree in Wireless Communication Systems (Distinction) from the University of Sheffield, United Kingdom, in 2015. He is currently pursuing his Ph.D. with the James Watt School of Engineering, University of Glasgow, United Kingdom. He is an Associate Fellow, Recognising Excellence in Teaching. His research interests include energy performance optimization of 5G and beyond heterogeneous cellular networks, radio resource management, cognitive radio, Unmanned Aerial vehicle (UAV) aided communications, Self-Organizing Networks (SON) and application of machine learning to wireless communications networks.
\end{IEEEbiographynophoto}

\begin{IEEEbiographynophoto}
{Iftikhar Ahmad} obtained his BS degree in Computer Science from the University of Malakand, Chakdara, Dir Lower, Pakistan, in 2014 and completed his MS in Software Engineering from COMSATS University, Islamabad, Pakistan, in 2019. He worked as a research assistant and a teacher assistant in the department of Computer Science at COMSATS University Islamabad, Pakistan. He also served as a lecturer at the Department of Computer Science and Information Technology at the University of Lahore Islamabad Campus and is currently pursuing his PhD with the James Watt School of Engineering at the University of Glasgow, United Kingdom. His research interests include Unmanned Aerial Vehicle (UAV) Assisted Next Generation Wireless Networks, 5G and Beyond Wireless Communication, Machine Learning for Wireless Communication, Software Project Management, Software Project Scheduling, Behavioural Software Engineering, and Global Software Engineering. He is a student member of the IEEE and a member of the IEEE Antenna and Propagation Society (APS).
\end{IEEEbiographynophoto}

\begin{IEEEbiographynophoto}{Kenechi G. Omeke} received a B. Eng. degree in Electronic Engineering from the University of Nigeria, Nsukka in 2012 and an MSc in Communication Engineering (Distinction) from the University of Manchester, UK in 2015. He is currently a PhD candidate at the James Watt School of Engineering, University of Glasgow UK, where he is working on the Internet-of-things (IoT) for underwater applications. His research interests are in IoT networking and applications, underwater communication, device-to-device communication, sensor networks and the applications of AI in wireless communication and networking. 
\end{IEEEbiographynophoto}

\begin{IEEEbiographynophoto}{Metin Ozturk} received the B.Sc. degree in electrical and electronics engineering from Eskisehir Osmangazi University, Turkey, in 2013, the M.Sc. degree in electronics and communication engineering from Ankara Yildirim Beyazit University, Turkey, in 2016, and the Ph.D. degree from the Communications, Sensing, and Imaging Research Group, James Watt School of Engineering, University of Glasgow, U.K., in 2020. He worked as a Research Assistant (from 2013 to 2016) and a Lecturer (from 2020 to 2021) with Ankara Yildirim Beyazit University, where he is currently an Assistant Professor. His research interests include intelligent networking for wireless communication networks, with a focus on energy efficiency, mobility management, and radio resource management in cellular networks.
\end{IEEEbiographynophoto}

\begin{IEEEbiographynophoto}{Cihat Ozturk} received the B.Tech. degree from Yildiz Technical University, Istanbul, Turkey, in 2011, and the M.S. degree in industrial engineering from the Kocaeli University, Kocaeli, Turkey, in 2014. He is currently pursuing the Ph.D. degree with a topic on network theory in the Marmara University, Istanbul. He is currently a Research Assistant with Ankara Yildirim Beyazit University (AYBU). His research interests include the location theory, network theory, optimization, and metaheuristics. 
\end{IEEEbiographynophoto}

\begin{IEEEbiographynophoto}
{Ali Makine Abdel-Salam} obtained his BSc degree in computer engineering from king Faisal University of Chad, N’Djamena, Chad in 2018. Currently continuing his MS degree in computer engineering at Ankara Yildirim Beyazit University, Ankara , Turkey. His research interests are focused on Internet of Things, machine learning, objects detection, and unmanned aerial vehicle (UAV).

\end{IEEEbiographynophoto}

\begin{IEEEbiographynophoto}{Michael S. Mollel} received his BSc degree in telecommunication engineering from the University of Dar es Salaam, Dar es Salaam, Tanzania, in 2011. The MSc and PhD degrees in information and communication engineering from Nelson, Mandela African Institution of Science and Technology, (NM-AIST) Arusha Tanzania in 2014 and 2021, respectively. He conducted research tenure in the Communications, Sensing, and Imaging Research Group, James Watt School of Engineering, University of Glasgow, U.K. 2018 - 2020. He received the 2019 Best Paper Award in IEEE WCNC 2019 Workshop. His research interests are focused on 5G and beyond 5G mobile networks, the Internet of Things, artificial intelligence, intelligent computer, wireless networks and computer vision. 
\end{IEEEbiographynophoto}

\begin{IEEEbiographynophoto}
{Qammer H. Abbasi} (SMIEEE, MIET, FRET, FRSA), Dr Abbasi is a Reader with the James Watt School of Engineering, University of Glasgow, U.K., deputy head for Communication Sensing and Imaging group. He has published 350+ leading international technical journal and peer reviewed conference papers and 10 books and received several recognitions for his research including URSI 2019 Young Scientist Awards, UK exceptional talent endorsement by Royal Academy of Engineering, Sensor 2021 Young Scientist Award , National talent pool award by Pakistan, International Young Scientist Award by NSFC China, National interest waiver by USA and 8 best paper awards. He is a committee member for IEEE APS Young professional, Sub-committee chair for IEEE YP Ambassador program,  IEEE 1906.1.1 standard on nano communication, IEEE APS/SC WG P145, IET Antenna and Propagation and healthcare network.
\end{IEEEbiographynophoto}

\begin{IEEEbiographynophoto}{Sajjad Hussain} (SM’17) is a Senior Lecturer in Electronics and Electrical Engineering at the University of Glasgow, UK. He has served previously at Electrical Engineering Department, Capital University of Science and Technology (CUST), Islamabad, Pakistan as Associate Professor. Sajjad Hussain did his masters in Wireless Communications in 2006 from Supelec, Gif-sur-Yvette and PhD in Signal Processing and Communications in 2009 from University of Rennes 1, Rennes, France. His research interests include 5G self-organizing networks, industrial wireless sensor networks and machine learning for wireless communications. Sajjad Hussain is a senior member IEEE and fellow Higher Education Academy.
\end{IEEEbiographynophoto}

\begin{IEEEbiographynophoto}
{Muhammad Ali Imran} (M'03--SM'12) received the M.Sc. (Hons.) and Ph.D. degrees from Imperial College London, U.K., in 2002 and 2007, respectively. He is Dean Glasgow College UESTC and a Professor of communication systems with the James Watt School of Engineering, University of Glasgow, U.K. He is an Affiliate Professor at the University of Oklahoma, USA, a Visiting Professor at the 5G Innovation Centre, University of Surrey, U.K and Adjunct Research Professor in Artificial Intelligence Research Center (AIRC) of Ajman University. He is leading research in University of Glasgow for Scotland 5G Center. He has over 18 years of combined academic and industry experience, working primarily in the research areas of cellular communication systems. He has been awarded 15 patents, has authored/co-authored over 400 journal and conference publications, and has been principal/co-principal investigator on over £6 million in sponsored research grants and contracts. He has supervised 40+ successful Ph.D. graduates. He has an award of excellence in recognition of his academic achievements, conferred by the President of Pakistan. He was also awarded the IEEE Comsoc’s Fred Ellersick Award 2014, the FEPS Learning and Teaching Award 2014, and the Sentinel of Science Award 2016. He was twice nominated for the Tony Jean’s Inspirational Teaching Award.
\end{IEEEbiographynophoto}

\end{document}